\documentclass[10pt,a4paper]{article}
\usepackage{advmaterstyle}
\usepackage{pifont}
\IfFileExists{ulem.sty}{\usepackage[normalem]{ulem}}{}

\DeclareFontFamily{OMS}{ntxtlf}{}
\DeclareFontShape{OMS}{ntxtlf}{m}{n}{<->ssub * cmsy/m/n}{}

\newcommand{\cmark}{\ding{51}}
\newcommand{\xmark}{\ding{55}}
\let\oldepsilon\epsilon \let\epsilon\varepsilon \let\varepsilon\oldepsilon
\renewcommand{\vec}{\boldsymbol}

\articletype{Research Article}%

\volume{00}
\startpage{1}

\begin{document}

\title{Theory of Spin-splitter Magnetoresistance in Altermagnets}

\author[1]{Tim Kokkeler \orcid{0000-0001-8681-3376}}
\author[2,3,4]{Vitaly N. Golovach \orcid{0000-0001-7457-171X}}
\author[2,3]{F. Sebastian Bergeret \orcid{0000-0001-6007-4878}}

\authormark{KOKKELER \textsc{et al.}}
\titlemark{Theory of Spin-splitter Magnetoresistance in Altermagnets}

\address[1]{\orgdiv{Department of Physics and Nanoscience Center}, \orgname{University of Jyv\"askyl\"a}, \orgaddress{\state{P.O. Box 35 (YFL), FI-40014}, \country{Finland}}}
\address[2]{\orgdiv{Centro de F\'isica de Materiales (CFM-MPC)}, \orgname{Centro Mixto CSIC-UPV/EHU}, \orgaddress{\state{20018 Donostia-San Sebasti\'an}, \country{Spain}}}
\address[3]{\orgdiv{Donostia International Physics Center (DIPC)}, \orgaddress{\state{20018 Donostia-San Sebasti\'an}, \country{Spain}}}
\address[4]{\orgdiv{IKERBASQUE, Basque Foundation for Science}, \orgaddress{\state{48013 Bilbao}, \country{Spain}}}

\corres{Tim Kokkeler \email{tim.h.kokkeler@jyu.fi}\\[0.5\baselineskip]
        Vitaly N. Golovach \email{vitaly.golovach@ehu.eus}\\[0.5\baselineskip]
        F. Sebastian Bergeret \email{fs.bergeret@csic.es}}

\abstract[Abstract]{%
We develop a theory of angular-dependent magnetoresistance (ADMR) in metallic altermagnets
coupled to ferromagnetic insulators and establish criteria that distinguish them from
conventional compensated magnets with spin-orbit coupling.
We show that, once its full set of angular dependencies is established,
the spin-splitter magnetoresistance
(SSMR)---recently reported in the bilayer geometry of
H.~Chen \textit{et al.} [\textit{Adv. Mater.} \textbf{37}, 2507764 (2025)]---would constitute
a smoking-gun signature of collinear $d$-wave altermagnetism in metallic systems.
Although SSMR has been regarded as a close analogue of spin Hall magnetoresistance (SMR),
we demonstrate that the two differ qualitatively in three key respects:
SSMR depends solely on the relative orientation between the ferromagnetic magnetization and
the altermagnetic Néel vector, yields a longitudinal ADMR response of opposite sign,
and features a direct proportionality between longitudinal and transverse ADMR signals, absent in SMR.
Building on these distinctions, we further show that the full angular dependence of the SSMR
provides a practical recipe to extract the direction of the Néel vector
from customary magnetoresistance measurements.
These results provide a clear route to unambiguously identify altermagnets in transport.
}


\maketitle

\section{Introduction}

Altermagnets are a recently identified class of collinear magnets that, in only a few years, have grown into one of the most active themes in contemporary magnetism \cite{sinova2022emerging,Jungwirth2025Alter,jungwirth2024altermagnets}. Following their theoretical prediction, a rapidly expanding body of work has mapped out their distinctive electronic and transport signatures \cite{zarzuela2024transport,leiviska2024anisotropy,cheng2024field,liu2024giant,kokkeler2025quantum,naka2025nonrelativistic,sigales2026spin,sun2023spin,das2023transport,gonzalez2021efficient,he2025crucial,herasymchuk2025electric,hallberg2025visualization,hodt2026phonon,heras2025interplay}, and experiments have begun to report evidence of altermagnetic order in a growing list of materials \cite{reimers2024direct,fedchenko2024observation,mencos2025direct,guo2024direct,chen2025spin,he2025evidence}. What makes these materials so appealing is that they combine defining features of both ferromagnets and antiferromagnets while remaining distinct from either. Although their magnetic structure is fully compensated and carries no net magnetization, their electronic bands are strongly spin split, with the splitting set by the crystal symmetry rather than by spin--orbit coupling \cite{smejkal2022beyond}. The resulting momentum-dependent spin splittings are typically large and, being of nonrelativistic origin, robust, so that efficient spin polarization and pronounced anisotropic transport \cite{gonzalez2021efficient} emerge in the complete absence of a macroscopic moment. Altermagnets thus provide a platform for spin transport that is intrinsically resilient to external magnetic fields \cite{sinova2022emerging,jungwirth2024altermagnets,Jungwirth2025Alter}---a property of clear interest for spintronic applications.

Several materials have been predicted to host altermagnetism~\cite{sinova2022emerging}.
Among those most relevant to the present work are metallic systems with $d$-wave
altermagnetic order. These include the controversial RuO$_2$, whose magnetic ground state
has been the subject of intense debate in recent years~\cite{li2025exploration,Choi2026exploring},
as well as Mn$_5$Si$_3$~\cite{reichlova2024observation,mencos2025direct},
KRu$_4$O$_8$~\cite{smejkal2022beyond}, doped (Cr,Fe)Sb$_2$~\cite{mazin2021prediction},
KV$_2$Se$_2$O~\cite{jiang2025metallic}, Fe$_2$Se$_2$O~\cite{wu2024valley}, and some
perovskites~\cite{naka2025altermagnetic}. In addition, many antiferromagnets have recently
been predicted to host $d$-wave altermagnetic states at their surfaces~\cite{lange2026emergent}.
Several of these systems remain theoretical predictions, and only a few altermagnets have
so far been experimentally confirmed. This highlights the need for robust experimental
probes capable of identifying altermagnetism, which is precisely what the spin-splitter
magnetoresistance proposed in the present work provides.

The defining symmetry of altermagnets sets them apart from conventional antiferromagnets in a way that has direct, measurable consequences. Whereas conventional antiferromagnets are invariant under the combined action of time reversal $\mathcal{T}$ with a translation $t$ or an inversion $\mathcal{P}$, altermagnets require an additional rotation $\mathcal{R}$; that is, they are invariant under $\mathcal{T}\mathcal{R}$, $\mathcal{T}\mathcal{R}t$, or $\mathcal{T}\mathcal{R}\mathcal{P}$ \cite{sinova2022emerging}. This distinction is mirrored directly in the electronic structure. Collinear antiferromagnets with $\mathcal{T}\mathcal{P}$ symmetry show no spin splitting at all, whereas noncollinear antiferromagnets that lack this symmetry can develop an odd-in-momentum splitting, as in $p$-wave magnets \cite{hayami2020spontaneous,hellenes2023exchange}. Altermagnets, by contrast, exhibit an even-in-momentum spin splitting. Because this splitting is nonrelativistic in origin it can be large, and it enables efficient charge--spin interconversion through the spin-splitter effect and its inverse \cite{gonzalez2021efficient,he2025crucial,herasymchuk2025electric,hallberg2025visualization,hodt2026phonon,kokkeler2025quantum,heras2025interplay,sigales2026spin}, as well as a number of other transport phenomena that have only recently begun to be explored \cite{gonzalez2024anisotropic,sun2025tunneling}.

A natural and experimentally mature way to probe such phenomena is through magnetoresistance (MR) measurements, which form a central class of spin-dependent transport probes in magnetic systems. By tying the electrical resistance to the magnetic configuration, to spin-dependent scattering, and to spin-orbit coupling, MR provides direct access to both magnetic order and spin-charge interconversion. The family of MR effects is broad, and includes anisotropic magnetoresistance (AMR) \cite{thomson1857xix}, giant magnetoresistance (GMR) \cite{baibich1988giant,binasch1989enhanced,camley1989theory}, tunnel magnetoresistance (TMR) \cite{julliere1975tunneling,liu2024giant}, Hanle magnetoresistance (HMR) \cite{dyakonov2007magnetoresistance,VelezHMR2016}, and spin Hall magnetoresistance (SMR) \cite{nakayama2013spin,chen2013theory,zhang2019theory,HahnSMR2013,althammer2013quantitative,isasa2014spin,meyer2014temperature,marmion2014temperature,kim2016spin,velez2019spinhall,uchida2015spin,gomez2020strong}; the latter two are regarded as hallmark signatures of the spin Hall effect (SHE) in heavy-metal systems.
\begin{figure*}[!t]
\centering
\begin{minipage}{0.55\textwidth}
    \centering
    \vspace{0pt}%
    \includegraphics[width = \textwidth]{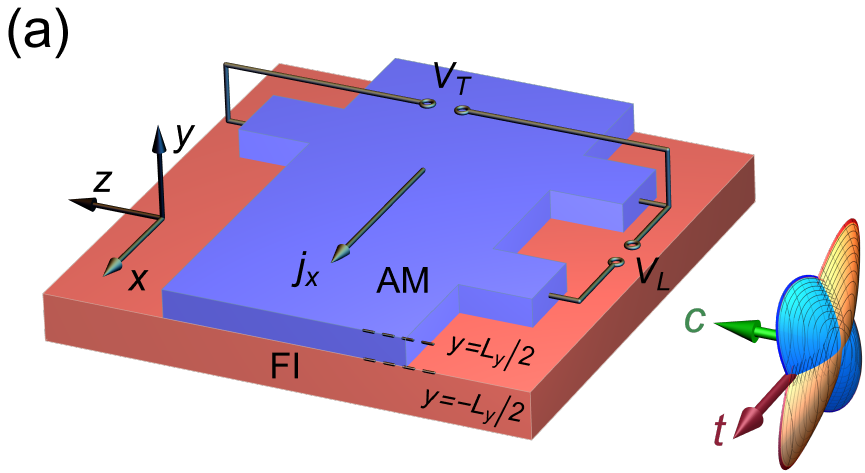}\\
\end{minipage}\hfill
\begin{minipage}{0.35\textwidth}
    \centering
    \vspace{0pt}%
    \includegraphics[width = \textwidth]{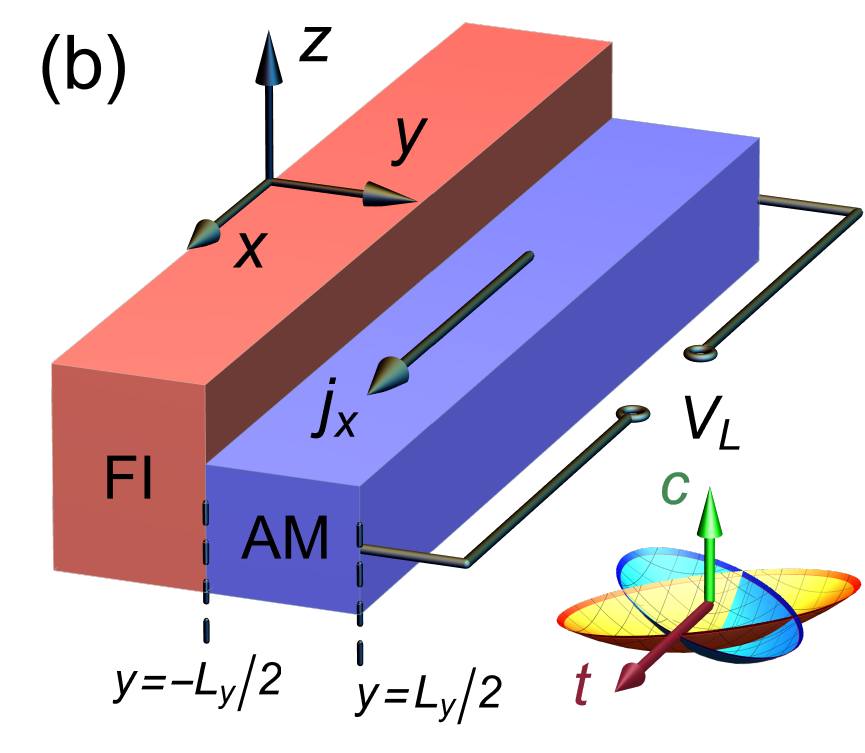}
\end{minipage}\\
\vspace{16pt}%
\begin{minipage}{0.58\textwidth}
    \centering
    \includegraphics[width = \textwidth]{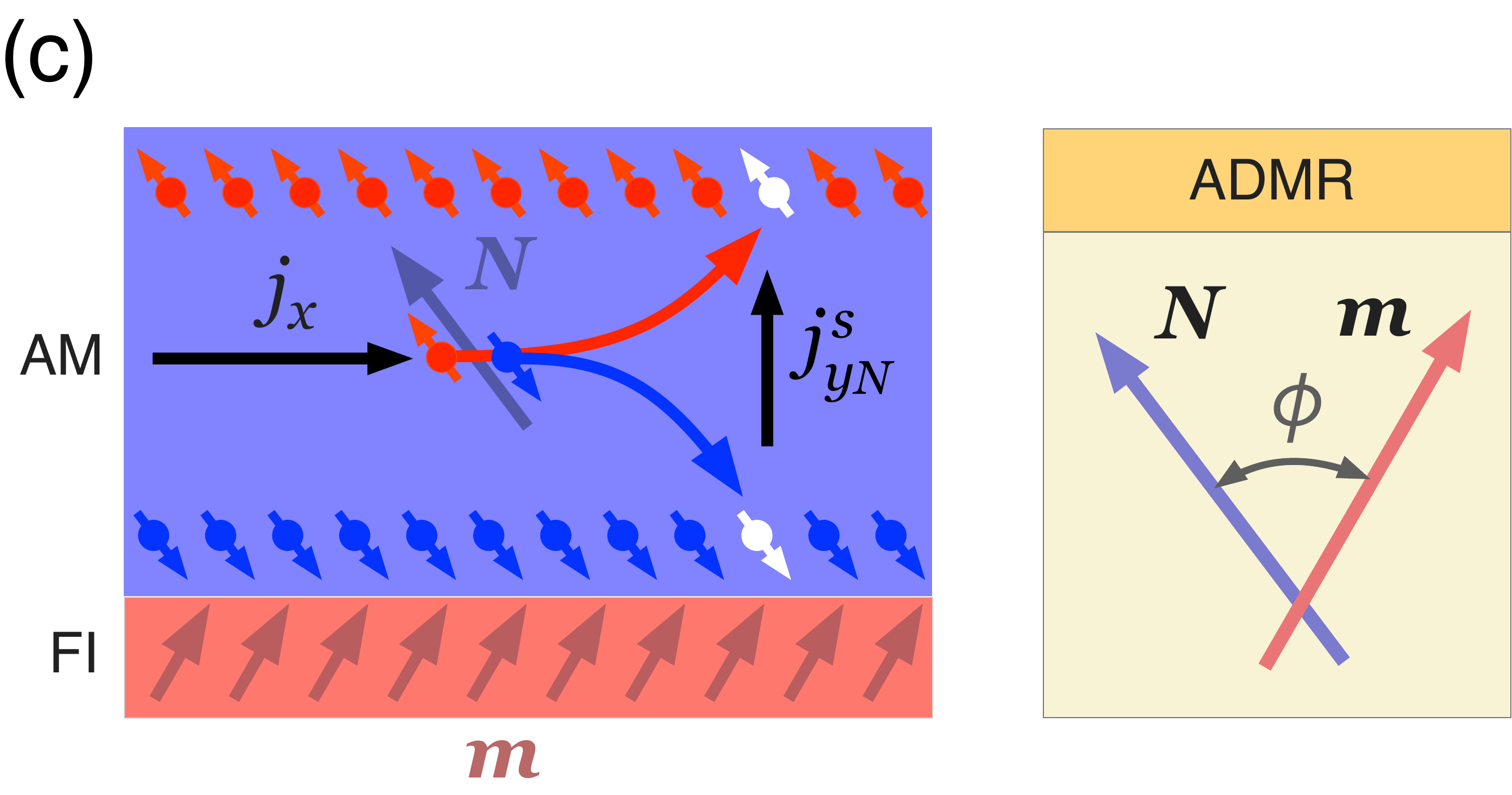}
\end{minipage}\hfill
\begin{minipage}{0.32\textwidth}
    \centering
    \includegraphics[width = \textwidth]{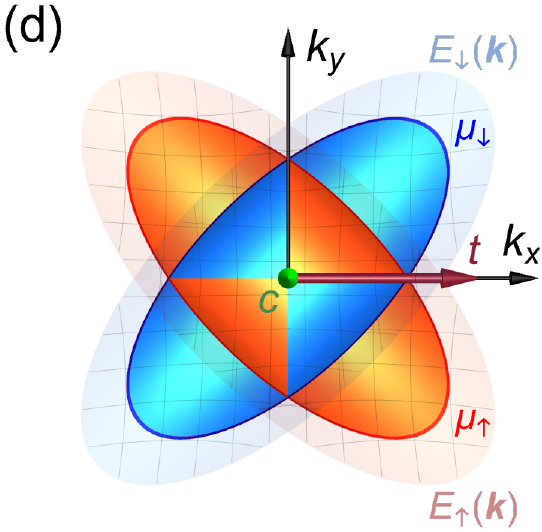}
\end{minipage}
    \caption{%
Device geometries and microscopic origin of the spin-splitter magnetoresistance (SSMR) in a collinear $d$-wave altermagnet (AM) coupled to a ferromagnetic insulator (FI) with magnetization direction $\bm{m}$.
(a) Hall-bar device with an AM thin film grown epitaxially on an FI substrate in a non-$c$-axis orientation.
Longitudinal ($V_L$) and transverse ($V_T$) voltages probe the SSMR response.
(b) Lateral-coupling geometry that isolates the purely longitudinal SSMR response.
This configuration provides a practical alternative to (a) when non-$c$-axis epitaxial growth of the AM is unavailable
(e.g., in strictly two-dimensional systems).
Insets in (a) and (b) show the altermagnetic ``flower'' orientation that maximizes the longitudinal response.
(c) Schematic illustration of spin accumulation at the AM interfaces.
An applied charge current density $j_x$ generates, via the spin-splitter effect (SSE),
a spin current density $j_{yN}^s$ polarized along the Néel vector $\bm{N}$.
The spin accumulation that builds up near the AM/FI interface is affected by interactions with the FI,
leading to an angular-dependent magnetoresistance (ADMR)
whose magnitude depends on the angle $\phi$ between $\bm{N}$ and $\bm{m}$ (inset).
(d) Top view of the band structure of a $d$-wave altermagnet:
the altermagnetic “flower” arises from the intersection of the spin-split bands
$E_{\uparrow}$ and $E_{\downarrow}$, occupied up to chemical potentials $\mu_{\uparrow}$ and $\mu_{\downarrow}$.
The spin-splitter vector $t_i$ and the crystal $c$-axis $c_i$ define the orientation of the spin-splitter
tensor $T_{ij}$  [Eq.~(\ref{eq:Tij})].
}
    \label{fig:SetupSSMR}
\end{figure*}

In this work we ask a simple but consequential question: can altermagnets be distinguished from conventional compensated magnets---in particular, from antiferromagnets with spin-orbit coupling---purely on the basis of MR measurements? To address it, we develop a theory of angular-dependent magnetoresistance (ADMR) in metallic $d$-wave altermagnets coupled to a magnetic insulator. This bilayer is closely related to the metallic altermagnet/ferromagnet structure that was recently explored experimentally in Ref.~\cite{chen2025spin}, and we find that the associated spin-splitter magnetoresistance (SSMR), a signature of which was reported in that work \cite{chen2025spin}, constitutes a genuine hallmark of altermagnetism---provided that its full set of angular dependencies is measured. We emphasize at the outset that, on the basis of the data currently available, we can neither confirm nor rule out that the MR corrections observed for the RuO$_2$ thin films of Ref.~\cite{chen2025spin} originate from intrinsic altermagnetism. What our theory does provide is a clear and practical route to establishing altermagnetic order, based on several distinctive features of the SSMR that we detail below.

These features emerge once SSMR is compared carefully with its spin--orbit-driven counterpart. SSMR was originally introduced in close analogy to SMR, with the spin Hall angle simply replaced by the spin-splitter mechanism; here, however, we show that the two differ qualitatively. As a nonrelativistic effect, SSMR is expected to produce larger signals than SMR and---more importantly for the purpose of identification---it carries distinct transport fingerprints. In particular, its longitudinal response has an angular dependence opposite to that of SMR, while its transverse response is directly proportional to the longitudinal correction, which is not the case for SMR. Taken together, these properties establish SSMR as a distinct class of ADMR, separate from spin--orbit-driven magnetoresistances such as SMR and HMR, and identify it as a powerful probe of altermagnetism. Building on this angular dependence, we further show that customary magnetoresistance measurements give direct access to the orientation of the Néel vector, providing a practical recipe to extract its direction from the measured ADMR.

\section{Model and transport equations}


We consider the setups shown in Fig.~\ref{fig:SetupSSMR}, consisting of a collinear altermagnet (AM) with Néel vector $N_a$ in contact with a ferromagnetic or ferrimagnetic insulator (FI) at $y=-L_y/2$.
To describe charge and spin dynamics, we employ the diffusive kinetic equations \cite{kokkeler2025quantum}:
\begin{align}
\partial_t \delta n + \partial_k j_k &= 0,\label{eq:ChargeCurrentconservatio}\\
\partial_t \delta S_a + \partial_k j^s_{ka} &= -\Gamma_{ab} \delta S_b,\label{eq:spindiffusionequation}
\end{align}
where $\delta n$ and $\delta \vec{S}$ denote the excess charge and spin densities,
with $j_k$ and $j^s_{ka}$ their associated currents,
and $\Gamma_{ab}$ is the spin-relaxation tensor. As diffusion equations, Eqs.~(\ref{eq:ChargeCurrentconservatio}) and (\ref{eq:spindiffusionequation}) are coarse-grained over the momentum relaxation time $\tau_p$ and therefore resolve only times longer than $\tau_p$.
In collinear altermagnets,
\begin{eqnarray}
\Gamma_{ab} &=& \tau_{s\perp}^{-1}(\delta_{ab}-N_aN_b)+\tau_{s\parallel}^{-1} N_a N_b,
\label{eq:Gammaab}
\end{eqnarray}
with $\tau_{s\parallel}$ and $\tau_{s\perp}$ the relaxation times for spin components parallel and perpendicular to the Néel vector. These times generally differ: the altermagnetic band splitting couples to transverse spins while leaving the longitudinal component largely unaffected.
Transverse spin components therefore undergo
efficient dephasing due to inhomogeneous broadening arising from the momentum dependence of the altermagnetic splitting $E_{\uparrow}(\bm{k})-E_{\downarrow}(\bm{k})$, where $\bm{k}$ is the electron wave vector.
For example, for a $d$-wave splitting $E_{\uparrow}(\bm{k})-E_{\downarrow}(\bm{k})=4\hbar Jk_xk_y/k_F^2$ \cite{maeda2025classification}
treated perturbatively
for electrons on a spherical Fermi surface $|\bm{k}|=k_F$ in 3D, the variance of the precession frequency is $\langle \delta\omega^2\rangle=(16/15)J^2$, yielding a dephasing time $\sim 1/J$, set by the band-splitting scale $J$. This regime corresponds to very clean systems, with momentum relaxation time $\tau_p\gtrsim 1/J$. Since the diffusive description resolves only times longer than $\tau_p$, a transverse dephasing time $\sim 1/J$ shorter than $\tau_p$ cannot be resolved and enters Eq.~(\ref{eq:spindiffusionequation}) as the limiting value $\tau_{s\perp}\approx 0$. In contrast, $\tau_{s\parallel}$ remains long ($\tau_{s\parallel}\gg \tau_p$), being limited by spin-orbit scattering from residual disorder and random magnetic disorder.

In the opposite, disordered limit ($\tau_p J\ll 1$),
the band splitting is no longer resolved between scattering events, and motional narrowing instead leads to
\begin{eqnarray}
\frac{1}{\tau_{s\perp}} &=& \frac{1}{\tau_{s\parallel}} +\alpha\tau_pJ^2,
\label{eq:ratesreldiffreg}
\end{eqnarray}
with $\alpha=16/15$ for the 3D model above and $\alpha=2$ in 2D. This regime bears some resemblance to the Dyakonov-Perel mechanism, although here the effect originates from fluctuations of the precession frequency without randomization of the precession axis \cite{bergeret2013singlet,vasiakin2025disorder,heras2025interplay}. We note that Eq.~(\ref{eq:ratesreldiffreg}) contains no interference terms between spin-orbit scattering and altermagnetic splitting fluctuations, owing to their different parity in $\bm{k}$.

For $d$-wave altermagnets, in the stationary limit the charge and spin current densities entering Eqs.~(\ref{eq:ChargeCurrentconservatio}) and (\ref{eq:spindiffusionequation}) read~\cite{kokkeler2025quantum}:
\begin{align}\label{eq:ChargeCurrent}
j_k &= \frac{\sigma}{e} \left(\partial_k \mu + \frac{1}{2} N_a T_{kj}\partial_j \mu_a^s\right)\;,
\end{align}
\begin{align}
j^s_{ka} &= -\frac{\hbar\sigma}{4e^2} \left(\partial_k \mu^s_a + 2N_a T_{kj}\partial_j\mu + \epsilon_{abc}N_b K_{kj}\partial_j \mu^s_c\right)\;, \label{eq:Spincurrent}
\end{align}
where $\sigma=2e^2\nu_0 D$ is the Drude conductivity, $e$ is the elementary charge, $\nu_0$ is the density of states per spin ($\nu_\uparrow=\nu_\downarrow\equiv \nu_0$),
$D$ is the diffusion constant at the Fermi level,
and $\epsilon_{abc}$ is the Levi-Civita symbol.
The mean chemical potential is defined as $\mu=(\mu_{\uparrow}+\mu_{\downarrow})/2$
and the spin bias as $\boldsymbol{\mu}^s=2\delta\boldsymbol{S}/(\hbar\nu_0)$, such that $\bm{N}\cdot\boldsymbol{\mu}^s=\mu_{\uparrow}-\mu_{\downarrow}$, where $\mu_{\uparrow}$ and $\mu_{\downarrow}$ denote the chemical potentials of the spin-up and spin-down bands.
Altermagnetism in Eqs.~(\ref{eq:ChargeCurrent}) and (\ref{eq:Spincurrent}) is encoded in two symmetric tensors: (i) the spin-splitter tensor $T_{jk}$ \cite{kitz1965symmetriegruppen,brinkman1966space,brinkman1966theory,litvin1974spin,litvin1977spin,liu2022spin,jiang2023enumeration,chen2024enumeration,xiao2023spin,shinohara2024algorithm,schiff2025crystallographic,feng2024superconducting,kokkeler2025quantum}, which couples charge and spin degrees of freedom, and (ii) the spin-swap tensor $K_{jk}$, analogous to the spin-swapping term in the SHE \cite{lifshits2009swapping}, which affects spin currents with polarization perpendicular to the Néel vector. In collinear systems, $K_{jk}$ does not contribute directly to the charge current~\cite{kokkeler2025quantum} and will be neglected in the following.

The spin-splitter tensor $T_{ij}$ is fixed by crystal symmetry.
For example, for a $d$-wave altermagnet with parabolic dispersion, it takes the form
\begin{eqnarray}
T_{ij} = \frac{2\hbar J}{E_F}\left(t_i\epsilon_{jkl}+t_j\epsilon_{ikl}\right)c_k t_l\;,
\label{eq:Tij}
\end{eqnarray}
where $E_F$ is the Fermi energy and $\bm{t}$ and $\bm{c}$
are unit vectors along the direction of maximal spin-splitter response
and the crystallographic $c$-axis, respectively [see Fig.~\ref{fig:SetupSSMR}(d)].
This tensor encodes the spin-splitter effect (SSE): spin-up and spin-down carriers are deflected anisotropically,
generating a transverse spin current polarized parallel to $\bm{N}$ [see Fig.~\ref{fig:SetupSSMR}(c)].
Although $\bm{N}$ can be reoriented by external fields or torques,
$\bm{t}$ and $\bm{c}$ are rigidly fixed by the lattice.
Consequently, the spin-splitter response depends strongly on the crystallographic orientation.

The spin accumulation vanishes when the interface 
normal is along the $c$-axis.
More generally, it also vanishes when the interface plane contains either of the vectors
$\bm{t}+\bm{c}\times\bm{t}$ or $\bm{t}-\bm{c}\times\bm{t}$.
In contrast, the accumulation is maximal when the charge current is aligned with one of the vectors
$\bm{t}$ or $\bm{c}\times\bm{t}$ and the interface normal is parallel to the remaining vector.

Figure~\ref{fig:SetupSSMR}(a) depicts the Hall-bar geometry with a FI at the bottom interface,
as typically used in SMR experiments.
Here the altermagnet is oriented such that its $d$-wave ``flower''
optimally generates spin accumulation at the top and bottom interfaces.
Notably, (101)-oriented epitaxial RuO$_2$ thin films were employed in Ref.~\cite{chen2025spin}
precisely to produce a misalignment between the altermagnetic ``flower'' and the film plane.
While such non-$c$-axis epitaxial growth is readily achieved in many rutile oxides via substrate templating,
it may not be feasible for all altermagnets---especially strictly two-dimensional systems.
In those cases, the lateral-coupling geometry shown in Fig.~\ref{fig:SetupSSMR}(b) provides a practical alternative.
The results presented below apply to both configurations.


We compute the longitudinal and transverse resistances as functions of the FI magnetization direction by solving the transport equations with appropriate boundary conditions. At the vacuum interface ($y=L_y/2$), the normal components of the charge and spin currents vanish. At the AM/FI interface ($y=-L_y/2$), the charge current is set to zero, while the spin current satisfies \cite{brataas2000finite,chen2013theory,dejene2015control,zhang2019theory}
\begin{align}
\frac{2e^2}{\hbar} \vec{j}_{y}^s &= G_i \vec{m}\times \vec{\mu}^s
+ G_r \vec{m}\times (\vec{m}\times\vec{\mu}^s)-|G_s|  \vec{\mu}^s\;, \label{eq:FIBC}
\end{align}
where $\bm{m}$ is the unit vector along the FI magnetization and the vector notation refers to spin space. As in SMR, the interface is parametrized by the spin-mixing conductance $G_{\uparrow\downarrow}=G_r+iG_i$ and the spin-sink conductance $G_s<0$. These parameters depend on temperature and magnetic field \cite{zhang2019theory} and are sensitive to the magnetic state of the FI and interfacial disorder.

We consider the linear response to a stationary, homogeneous charge-current density $j_x$ in the geometries of
Figs.~\ref{fig:SetupSSMR}(a) and (b).
We retain the transport coordinate frame shown in the figures while allowing an arbitrary crystal-lattice orientation,
further considering a generic spin-splitter tensor $T_{ij}$
which leads to spin accumulation at the AM interfaces via the SSE [see Fig.~\ref{fig:SetupSSMR}(c)].
Translational invariance requires the chemical potential and spin bias to take the form
$\mu=-Ex+\delta\mu(y)$ and $\partial_x\mu_k^s=0$,
where $E$ is the uniform $x$-component of the electric field.
With this ansatz, the continuity equations
(\ref{eq:ChargeCurrentconservatio}) and (\ref{eq:spindiffusionequation}), together with
 the currents in Eqs.~(\ref{eq:ChargeCurrent}) and (\ref{eq:Spincurrent}),
reduce to a one-dimensional problem for $\delta\mu(y)$ and $\mu_k^s(y)$.
From the solutions to this problem, the charge current can be calculated with the help of Eq. (\ref{eq:ChargeCurrent}).
Averaging Eq. (\ref{eq:ChargeCurrent})
over the film thickness yields
$\bar{j}_x=\sigma E + \delta\bar{j}$, with
\begin{align}
    \delta\bar{j} = -\frac{\sigma T_{xy}}{2eL_y}\vec{N}\cdot\left[\vec{\mu}^s\left(\frac{L_y}{2}\right)-\vec{\mu}^s\left(-\frac{L_y}{2}\right)\right]\;.\label{eq:Jx}
\end{align}
This expression closely resembles the SMR,
except that the role of the spin Hall angle is played by the tensor component $T_{xy}$
and the spin accumulation enters solely through its projection onto $\bm{N}$.
The same projection governs the transverse current $j_z$, which obeys the simple proportionality
\begin{align}
\bar{j}_z = \frac{T_{yz}}{T_{xy}} \delta\bar{j}\;.
\label{eq:Jz}
\end{align}
This relation stands in stark contrast to the SHE (and related phenomena such as SMR and HMR),
where the longitudinal and transverse responses depend on independent components of the spin accumulation
($\mu_z$ and $\mu_x$, respectively)
rather than on the single combination $\bm{N}\cdot\bm{\mu}^s$.
Consequently, the ADMR of the SSMR differs markedly from that of SMR %
when both the longitudinal and transverse signals are measured.

\section{Results}

For the longitudinal ($\rho_L$) and transverse ($\rho_T$) resistivities, we obtain (see Appendix~\ref{sec:SSMRderivation})
\begin{eqnarray}
\rho_L = \sigma^{-1} + \Delta \rho_L
\quad\text{and}\quad
\rho_T= \frac{T_{yz}}{T_{xy}} \Delta \rho_L\;,
\label{eq:rhoLrhoT}
\end{eqnarray}
where
\begin{strip}
\begin{align}\label{eq:Deltarhogeneral}
\Delta \rho_L = \frac{2T_{xy}^2l_{s_{\parallel}}}{\sigma L_y}\tanh\left(\frac{L_y}{2l_{s\parallel}}\right)
\frac{\sigma+|G_s|l_{s\parallel}\coth\left(\frac{L_y}{2l_{s\parallel}}\right)
+\bar{G}\left[l_{s\parallel}\coth\left(\frac{L_y}{2l_{s\parallel}}\right)
-2l_{s\perp}\coth\left(\frac{L_y}{l_{s\perp}}\right)\right]\sin^2\phi}
{\sigma+2|G_s|l_{s\parallel}\coth\left(\frac{L_y}{l_{s\parallel}}\right)
+2\bar{G}\left[l_{s\parallel}\coth\left(\frac{L_y}{l_{s\parallel}}\right)
-l_{s\perp}\coth\left(\frac{L_y}{l_{s\perp}}\right)\right]\sin^2\phi}\;,
\end{align}
\end{strip}
with spin-relaxation lengths $l_{s\parallel,\perp}=\sqrt{D\tau_{s\parallel,\perp}}$, angle $\phi$ between $\bm{N}$ and $\bm{m}$, and effective spin-mixing conductance
\begin{align}
\bar{G} =
\mathrm{Re}\frac{G_{\uparrow\downarrow}}
{1+\frac{2l_{s\perp}}{\sigma}\left(|G_s|+G_{\uparrow\downarrow}\right)
\coth\left(\frac{L_y}{l_{s\perp}}\right)}\;.
\label{eq:barGeff}
\end{align}
Equations~(\ref{eq:rhoLrhoT})-(\ref{eq:barGeff}) constitute the central result of this work. 
They describe the ADMR of the AM/FI structures in Fig.~\ref{fig:SetupSSMR} 
and establish the theoretical framework for future SSMR experiments.

Quite remarkably, the imaginary part of the spin-mixing conductance ($G_i$)---a quantity
proportional to the interfacial exchange field induced by the FI in the AM---drops
out of the results in Eqs.~(\ref{eq:rhoLrhoT})--(\ref{eq:barGeff}) to first order.
The effect of both $G_r$ and $G_i$ on the MR corrections is captured compactly by the
effective spin-mixing conductance $\bar{G}$ in Eq.~(\ref{eq:barGeff}); since $\bar{G}$
is an even function of $G_i$, the leading correction is quadratic, and the transverse
resistivity $\rho_T$ acquires no contribution of the anomalous-Hall type.
This is in stark contrast with conventional SMR, for which effects directly proportional
to the interfacial exchange field are observed in high-quality samples such as
Pt/EuS~\cite{gomez2020strong} and Pt/Gd$_3$Ga$_5$O$_{12}$ (GGG)~\cite{Oyanagi2021para}.

\subsection{The non-chiral sign reversal}

For $l_{s\parallel} = l_{s\perp}$,
the resulting ADMR is purely harmonic:
\begin{align}\label{eq:Longitudinal}
\rho_L &= \sigma^{-1} + \Delta\rho_{0}+\Delta\rho_{1}\left[1-(\vec{m}\cdot\vec{N})^{2}\right]\;,\\
\rho_{T} &= \frac{T_{yz}}{T_{xy}}\left\{
     \Delta\rho_{0}+\Delta\rho_{1}\left[1-(\vec{m}\cdot\vec{N})^{2}\right]\right\}\;, \label{eq:rhoTiso}
\end{align}
where
\begin{align}\label{eq:Longitudinalrho0iso}
\Delta\rho_{0} &=  \frac{2 T_{xy}^{2}  l_s}{\sigma L_y }
    \frac{ \sigma \tanh{\left(\frac{L_{y}}{2l_{s}}\right)} + |G_s|l_s}
    {\sigma + 2|G_s|l_s\coth\left(\frac{L_y}{l_s}\right)}\;,\\
\Delta\rho_{1} &=
    -\frac{2T_{xy}^{2} l_s}{\sigma L_y}
    \frac{\bar{G}l_s \tanh^2\left({\frac{L_{y}}{2l_{s}}}\right)}{\sigma + 2|G_s| l_{s}\coth\left(\frac{L_y}{l_{s}}\right)}\label{eq:Longitudinalrho1iso}\;.
\end{align}
This functional form is identical to that of SMR,
except that the corrections
$\Delta\rho_0$ and $\Delta\rho_1$
in Eqs.~(\ref{eq:Longitudinalrho0iso})
and~(\ref{eq:Longitudinalrho1iso}) carry
the opposite sign relative to similar terms in conventional SMR~\cite{chen2013theory,zhang2019theory}.
This sign reversal is a direct hallmark of
the non-chiral nature of the SSE,
in contrast to the chiral nature of the SHE.

In the SHE, successive diffusive deflections of electrons exhibit a tendency to wind,
analogous to the action of a fictitious Lorentz force whose direction is spin-dependent owing to spin-orbit coupling.
By contrast, the SSE arises solely from the anisotropy of the spin-split band structure:
electrons preferentially drift along directions of maximum group velocity $\propto\partial E/\partial\bm{k}$,
so the deflection mechanism carries no intrinsic chirality.
This is directly reflected in the symmetric spin-splitter tensor $T_{ij}$.

\begin{figure}[!t]
    \centering
\begin{minipage}{0.6\textwidth}
\vspace{0pt}%
\begin{minipage}{0.38\columnwidth}
    \centering
    \vspace{0pt}%
    \includegraphics[width = \textwidth]{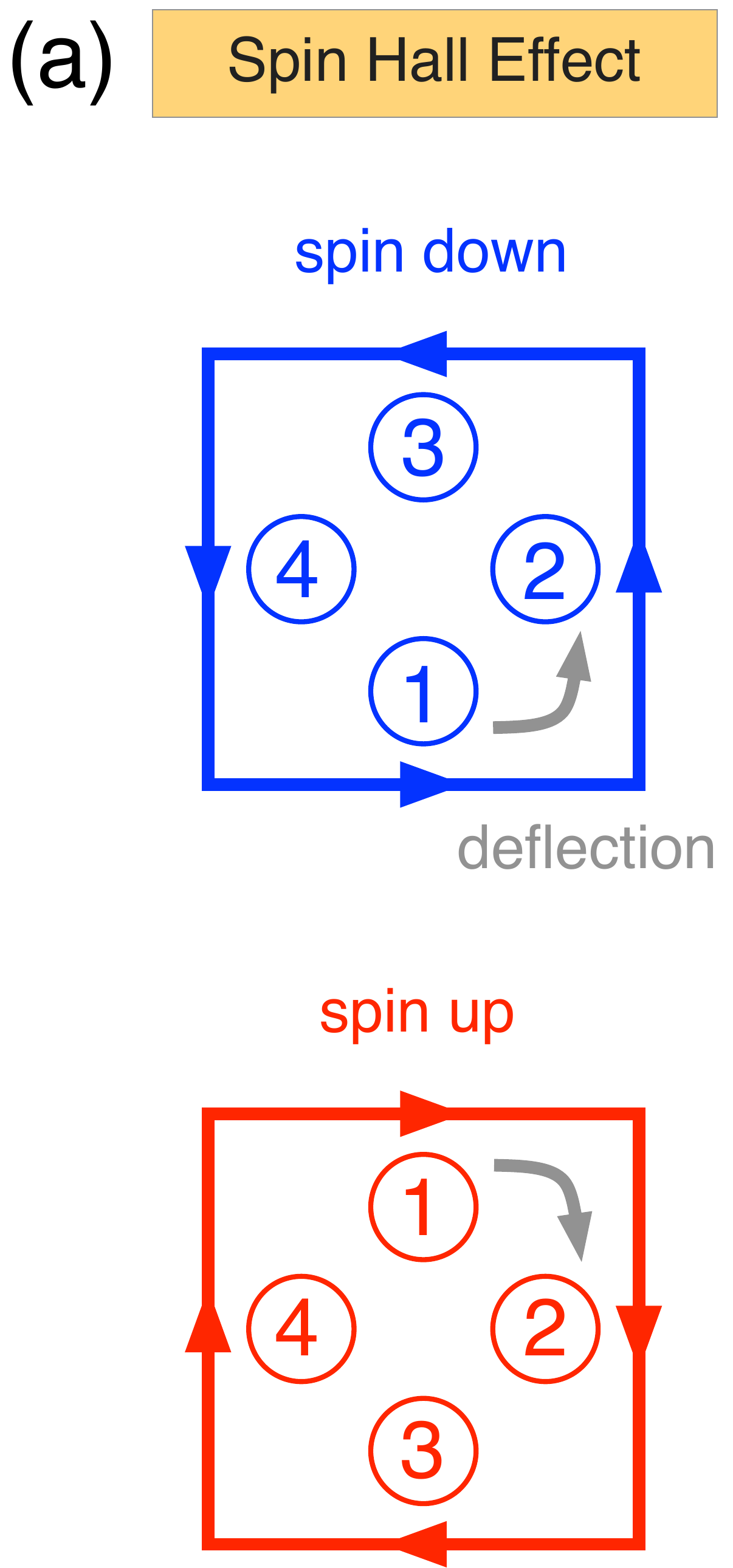}
\end{minipage}\hfill
\begin{minipage}{0.54\columnwidth}
    \centering
    \vspace{6pt}%
    \includegraphics[width = \textwidth]{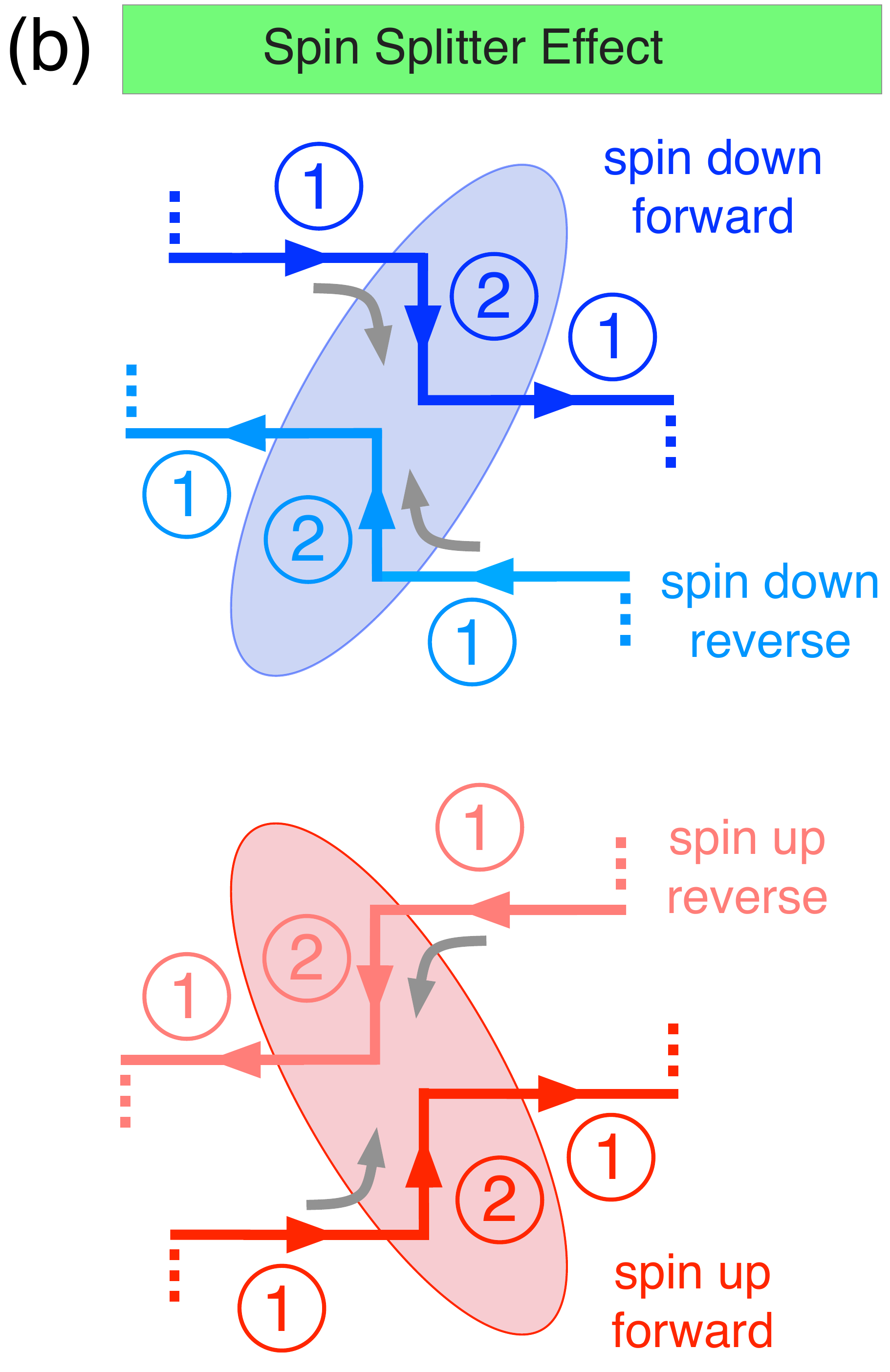}
\end{minipage}\\
\vspace{12pt}%
\begin{minipage}{0.48\columnwidth}
    \centering
    \vspace{0pt}%
    \includegraphics[width = \textwidth]{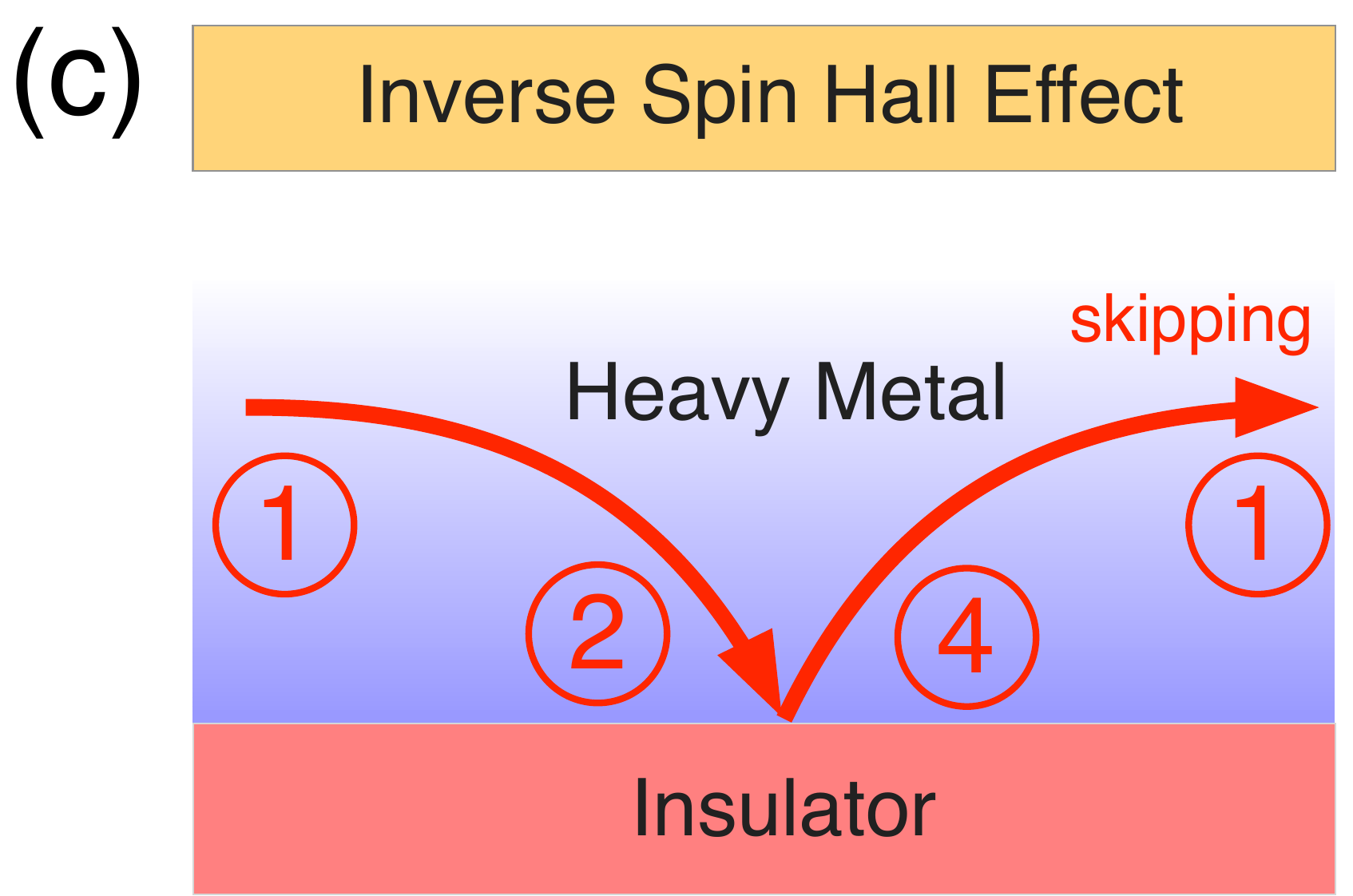}
\end{minipage}\hfill
\begin{minipage}{0.48\columnwidth}
    \centering
    \vspace{0pt}%
    \includegraphics[width = \textwidth]{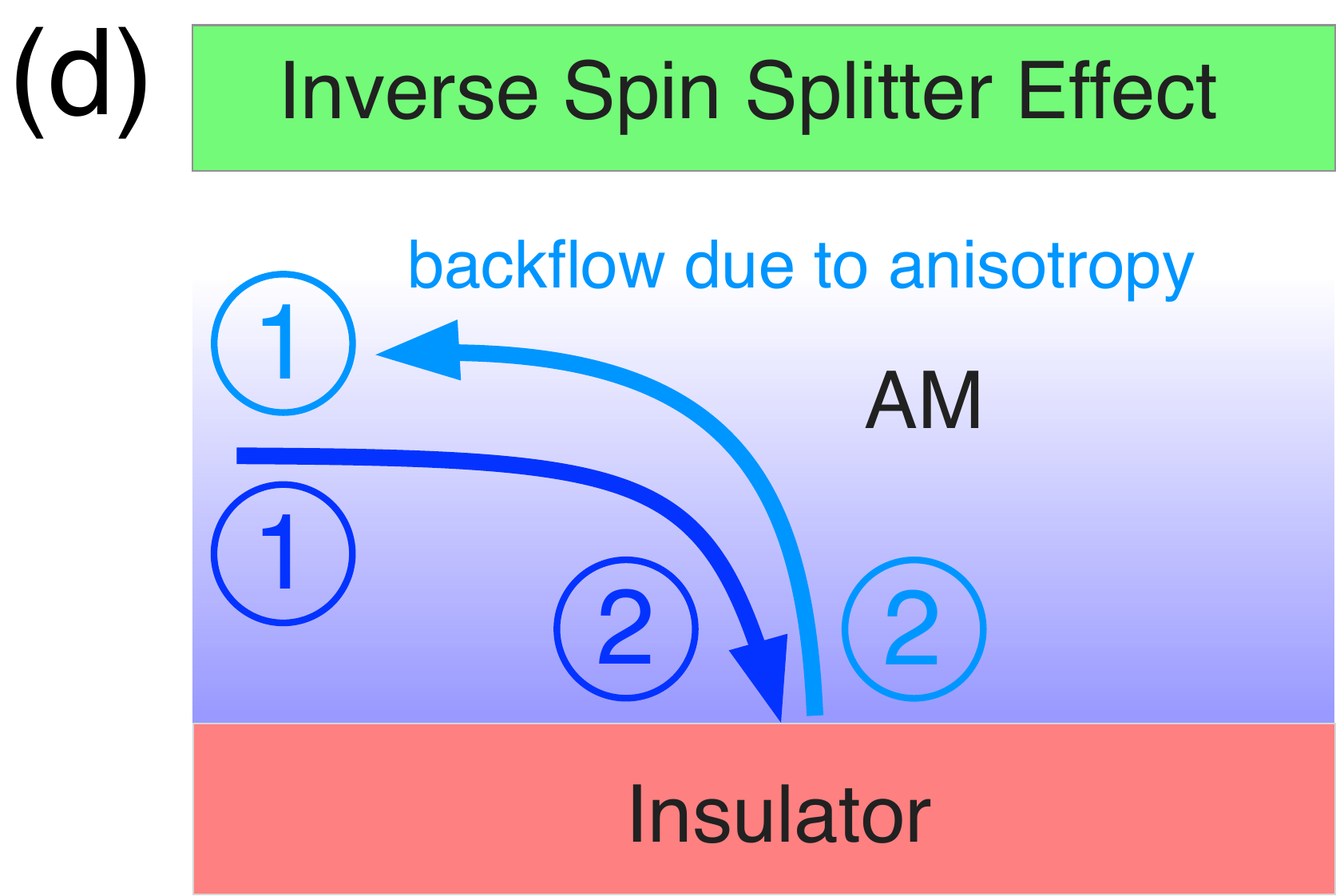}
\end{minipage}\\
\vspace{12pt}%
\begin{minipage}{0.48\columnwidth}
    \centering
    \vspace{0pt}%
    \includegraphics[width = \textwidth]{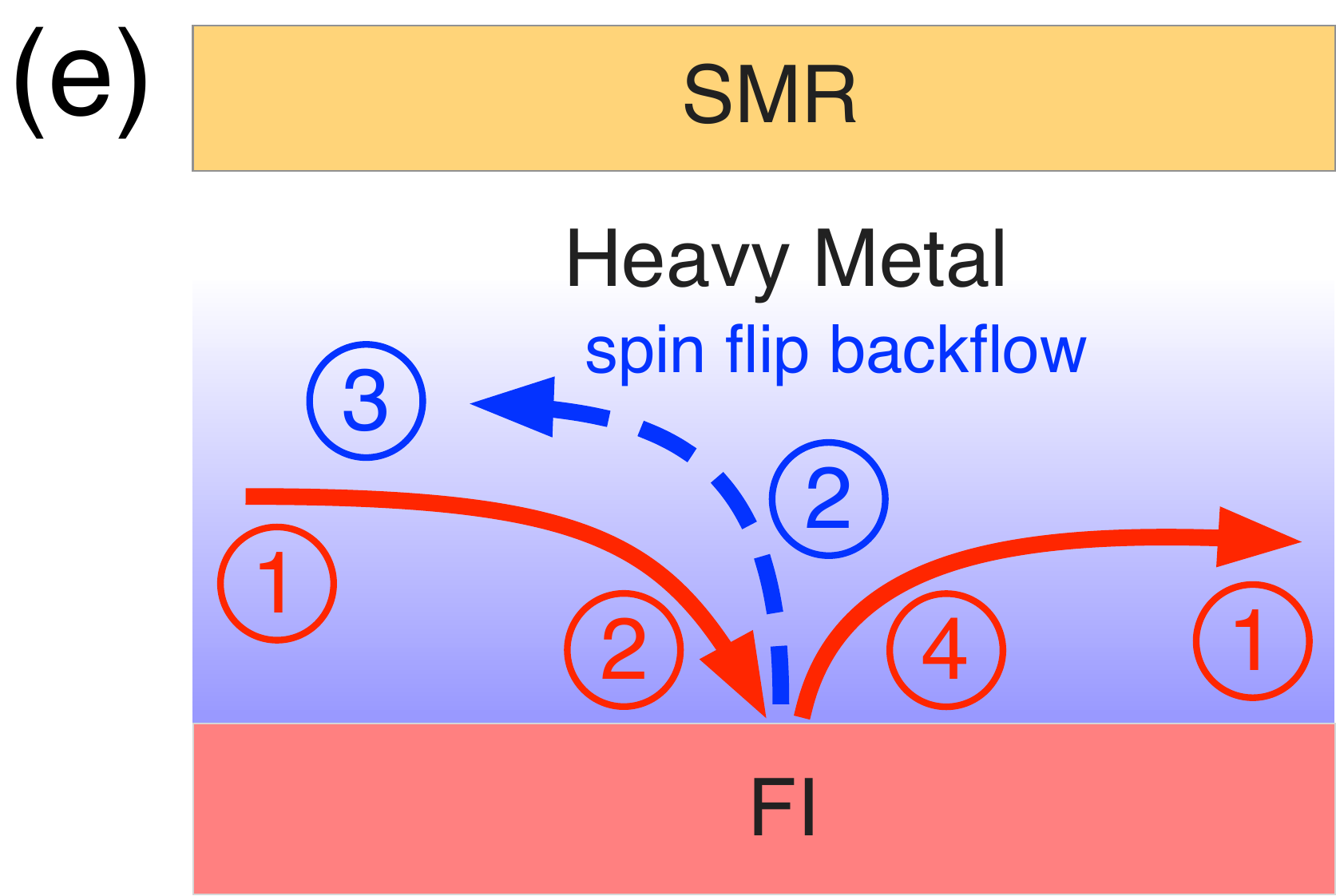}
\end{minipage}\hfill
\begin{minipage}{0.48\columnwidth}
    \centering
    \vspace{0pt}%
    \includegraphics[width = \textwidth]{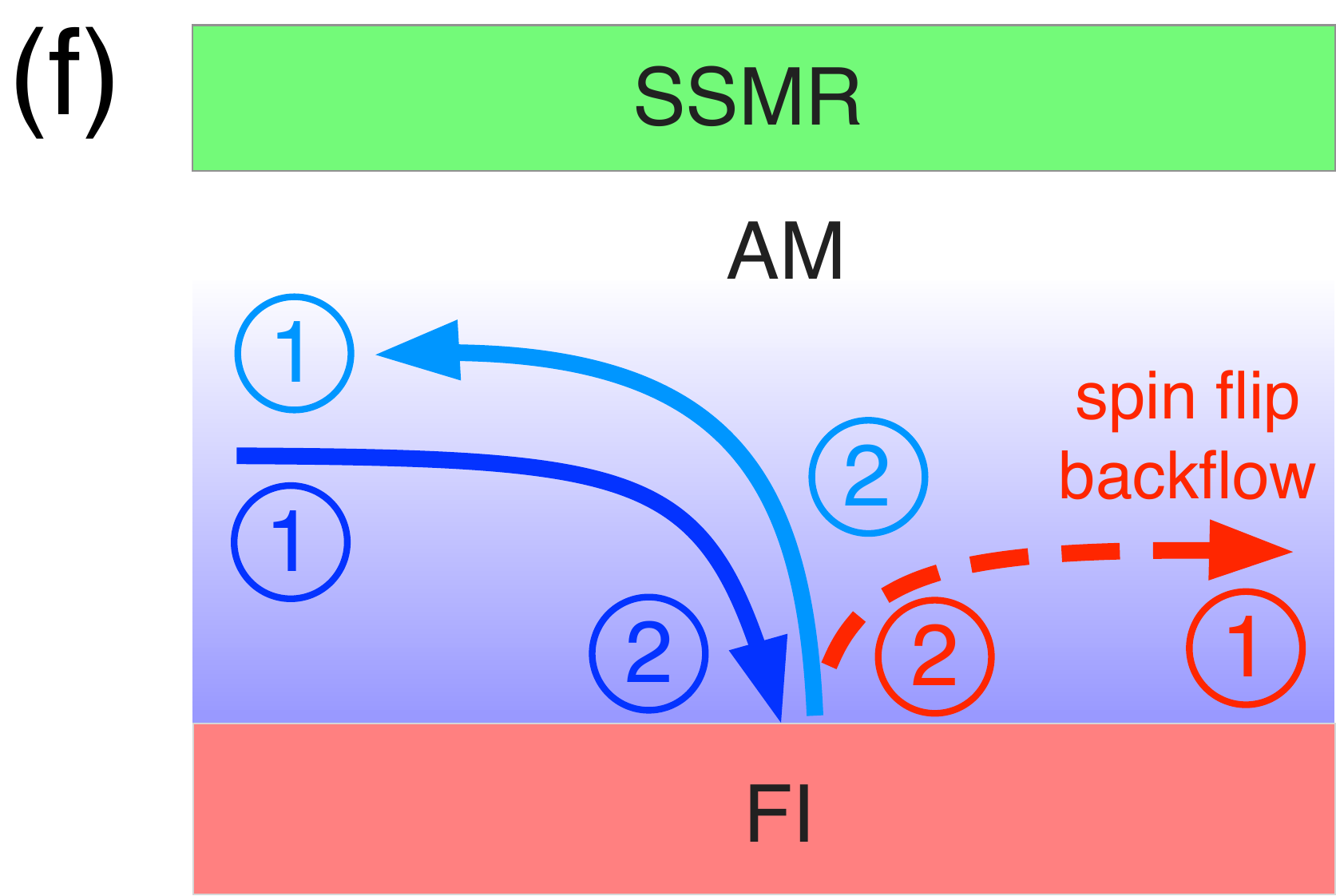}
\end{minipage}
\end{minipage}
\caption{
Chiral versus non-chiral deflection diagrams for the spin Hall effect (SHE) in a heavy metal and the spin-splitter effect (SSE) in a $d$-wave altermagnet.
(a)~Successive deflections in the SHE form closed square loops
with opposite chiralities for the two spin species,
reflecting the intrinsic handedness of the spin–orbit interaction.
(b)~In the SSE, consecutive deflections form an open staircase trajectory with two-fold periodicity
(\textcircled{1}$  \to  $\textcircled{2}$  \to  $\textcircled{1}$  \to\dots  $),
dictated by the anisotropic spin-split Fermi surface (shaded ellipse).
Forward and reverse sequences never intersect.
(c,d)~Inverse effects driven by spin accumulation at an interface.
(c)~The inverse SHE yields a skipping-orbit-like
backflow consistent with the chiral Lorentz-like force.
(d)~The inverse SSE produces a backflow governed solely by band-structure anisotropy.
(e,f)~SMR and SSMR arise when the inverse processes are modulated
by spin scattering at the FI interface, generating additional spin-flip backflows (dashed arrows).
}
    \label{fig:DeflecDiagram}
\end{figure}

Figure~\ref{fig:DeflecDiagram} illustrates this fundamental difference through deflection diagrams.
For the SHE [panel~(a)], successive deflections close after four steps
and possess opposite chiralities for spin-up and spin-down electrons.
In the SSE [panel~(b)], the deflections of both spin species
follow an open staircase trajectory with two-fold periodicity,
dictated by the anisotropic spin-split Fermi surface.
Consequently, the inverse SHE and the inverse SSE produce charge currents
in opposite directions [panels~(c) and~(d)].
The SMR (SSMR) arises when the FI interface modulates the inverse SHE (SSE)
via spin-dependent scattering,
generating additional backflows of electrons
with flipped spins (dashed arrows in panels~(e) and~(f)).
These backflow currents flow in the opposite (same) direction
as the bulk current for SMR (SSMR) and thus increase (decrease) the resistivity.

\subsection{Extracting the Néel vector from ADMR}

We summarize our results deep in the diffusive regime,
where the parallel and perpendicular spin diffusion lengths
are equal ($l_{s\parallel}=l_{s\perp}$), in Fig.~\ref{fig:AngleDependence},
with the left (right) column showing the ADMR of a collinear $d$-wave altermagnet
(conventional SMR).
In the upper panels, the full angular dependence is illustrated by spherical plots in which the radial distance
is proportional to $\rho_L$
(after subtraction of a constant to highlight the angular variation) and the direction corresponds to $\bm{m}$.
The opposite sign of the correction $\Delta\rho_1$ produces qualitatively different shapes: a peanut-shaped profile for SSMR
and a doughnut-shaped profile for SMR.

We consider line cuts in the standard $\alpha$,
$\beta$, and $\gamma$ rotation planes of $\bm{m}$
relative to the Hall-bar geometry,
as commonly employed in SMR studies~\cite{nakayama2013spin}.
The middle panels of Fig.~\ref{fig:AngleDependence} show these cuts for the longitudinal resistivity change $\Delta\rho_L$,
while the lower panels display the corresponding cuts for the transverse change $\Delta\rho_T$.

In SMR the symmetry axis of the ADMR
(with respect to rotations of $\bm{m}$) is set by
$\bm{n}\times\bm{j}$, where $\bm{n}$ is the interface normal
and $\bm{j}$ is the charge-current direction.
In SSMR this role is played by the Néel vector $\bm{N}$.
Because $\bm{N}$ can be oriented generically
(and can be controlled, for example, by magnetic torques),
we consider the representative orientation shown
in the top-left panel of Fig.~\ref{fig:AngleDependence}.
As a result, the SSMR line cuts not only reverse sign relative to SMR but also acquire shifts that reflect the specific orientation of $\bm{N}$.
These characteristic shifts furnish a practical recipe for extracting the direction of $\bm{N}$ (up to a sign)
from experimental ADMR line cuts.

Since $\bm{N}$ and $-\bm{N}$ are indistinguishable in SSMR,
we adopt the convention $N_Z>0$ in the coordinate frame of the figure.
We identify the angles $\alpha_N$, $\beta_N$,
and $\gamma_N$ of $\bm{N}$ with the positions of
the maxima in the ADMR line cuts.
The maxima for $\beta_N$ and $\gamma_N$
are chosen so that these angles lie in the interval
$[-\pi/2,\pi/2]$, consistent with $N_Z>0$.
For $\alpha_N$ we select the maximum that satisfies the auxiliary relation
\begin{align}
\tan \alpha_N \, \tan \gamma_N +\tan \beta_N=0\;.
\end{align}
which follows from the redundancy of the three-angle parametrization of the ADMR and provides a consistency check that helps exclude spurious mechanisms.

Once $\alpha_N$, $\beta_N$, and $\gamma_N$
have been determined, the components of $\bm{N}$
are obtained from
\begin{align}
N_Z =& \frac{\left|\sin\alpha_N\right|}{\sqrt{\sin^2\alpha_N+\tan^2\beta_N}}\;,\nonumber\\
N_Y =& -N_Z\tan\beta_N\;,\nonumber\\
N_X =& N_Z\tan\gamma_N\;.
\end{align}

We remark that applying the identical extraction procedure to conventional SMR would require identifying the minima rather than the maxima of the ADMR curves when determining the angles of
$\bm{n}\times\bm{j}$.
This reversal originates from the chiral minus sign characteristic of the SHE.
Consequently, an unaccounted $\pi/2$-shift
can appear when the intuition developed for SMR
is transferred directly to SSMR measurements~\cite{chen2025spin,he2025evidence}.

\section{Discussion}

In the specific case of RuO$_2$,
the experiment of Refs.~\cite{chen2025spin,he2025evidence} reports ADMR traces
in the $\beta$-plane that exhibit precisely such phase
shifts relative to conventional SMR reference curves.
For decades RuO$_2$ was regarded as a Pauli paramagnet
with no local moments on the Ru sites~\cite{Mukuda1999,li2025exploration,Choi2026exploring}.
More recently it has been proposed as a prototypical
altermagnet on the basis of density-functional calculations predicting momentum-dependent spin splitting~\cite{Ahn2019,Hayami2019,Smejkal2020,gonzalez2021efficient,smejkal2022beyond,sinova2022emerging}
and of early neutron diffraction and X-ray scattering reports of a
weak collinear Néel order~\cite{Berlijn2017,Zhu2019}.
These results were subsequently supported by transport measurements~\cite{Feng2022} and direct observation of spin-split
bands~\cite{Lin2024Observation,fedchenko2024observation}.
However, a growing body of more recent and systematic
measurements---spanning muon spin rotation/relaxation ($\mu$SR)~\cite{Hiraishi2024,Kessler2024},
neutron diffraction~\cite{Kiefer2025},
ARPES~\cite{Liu2024,Osumi2026}, magnetometry~\cite{Qian2026}, and transport~\cite{Plouff2025,PengXin2025,Wenzel2025,WuZheyu2025,Wang2026absence}---finds
no evidence of long-range magnetic order,
indicating that bulk RuO$_2$ is a conventional
Pauli paramagnet to high accuracy.
The apparent altermagnetic signatures reported in numerous heterostructure studies~\cite{Bai2022torque,Karube2022,Bai2023,guo2024direct,chen2025spin,he2025evidence,LiZhuoyi2025,ZhangYichi2025,Noh2025} may therefore arise predominantly from interfacial
spin-orbit effects
(e.g., interface-generated spin currents as reported in Ref.~\cite{Akashdeep2025} or surface states as observed in Ref.~\cite{Osumi2026})
or from strain- and defect-induced magnetism~\cite{Jeong2025metallicity,Akashdeep2026surface,Smolyanyuk2024,Ho2025symmetry,Jeong2026emergence}
that is absent in high-quality bulk crystals.
While our theory of SSMR assumes an intrinsic altermagnetic ground state, the present experimental situation in RuO$_2$
suggests
that interfacial contributions may play a significant role in the observed magnetoresistance~\cite{Akashdeep2025}.
Consequently, a direct one-to-one mapping between the SSMR signatures predicted here and existing RuO$_2$ data is not straightforward; careful separation of bulk altermagnetic and interfacial mechanisms remains essential.

We also remark that apparent ``negative SMR'' signals have been
reported in heavy-metal / antiferromagnetic-insulator bilayers (e.g., Pt on NiO
or $\alpha$-Fe$_2$O$_3$)~\cite{shang2016effect,hou2017tunable,
hoogeboom2017negative,baldrati2018full,cheng2019anisotropic,fischer2020large,wu2025control}.
In those systems the sign reversal (or negative-looking response) arises primarily when the magnetoresistance is plotted against the applied magnetic field without correcting for the actual orientation of the Néel vector $\bm{N}$ of the insulator.
In easy-plane antiferromagnets such as NiO the moments cant or rotate within the plane;
in easy-axis materials such as $\alpha$-Fe$_2$O$_3$ a spin-flop transition occurs at sufficiently high fields.
In both cases $\bm{N}$ tends to align perpendicular to the applied field, producing an effective sign change relative to the conventional positive SMR of a heavy-metal/ferromagnet-insulator bilayer.
Because this behavior is dictated by the magnetic configuration evolution of the insulating antiferromagnet,
this apparent ``negative SMR'' is a property of the magnetic \emph{insulator} rather than of the adjacent metal.
It is therefore straightforward to separate from the intrinsic sign reversal of SSMR.
In practice the two are easily told apart by routine control experiments: the apparent ``negative SMR''
disappears once the magnetoresistance is referenced to the actual orientation of $\bm{N}$ rather than to the applied field,
and it changes when the same metal is paired with a different insulator,
whereas the SSMR sign reversal originates from the non-chiral spin-splitter tensor
of the altermagnetic \emph{metal} and persists even for a fixed Néel-vector orientation.
Since the insulator and its magnetic anisotropy are known from the outset,
the two mechanisms can be distinguished unambiguously.

\begin{figure}[!t]
    \centering
\begin{minipage}{0.8\textwidth}
    \includegraphics[width = 0.48\columnwidth]{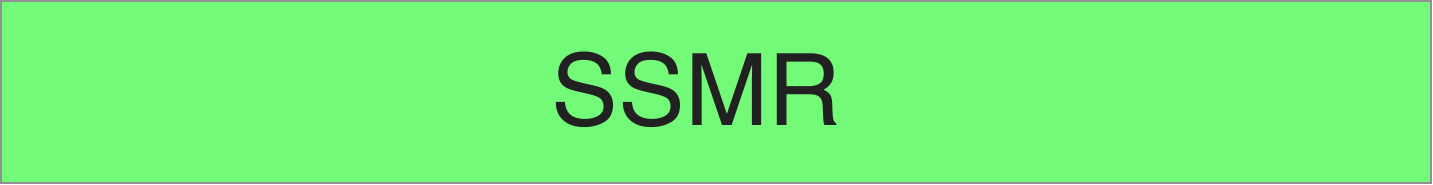}
    \includegraphics[width = 0.48\columnwidth]{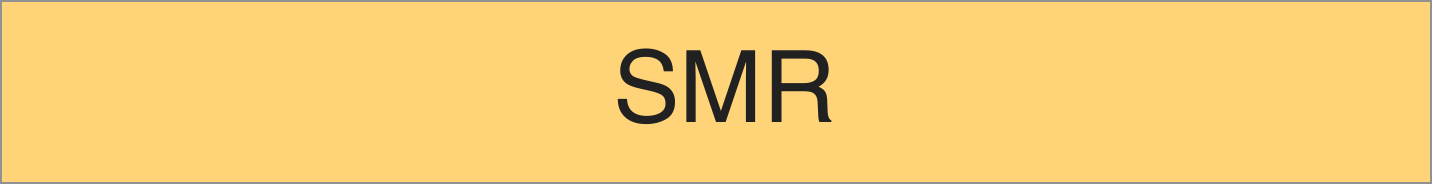}
    \vspace{0pt}%
\begin{minipage}{0.48\columnwidth}
    \centering
    \vspace{0pt}%
    \includegraphics[width = \textwidth]{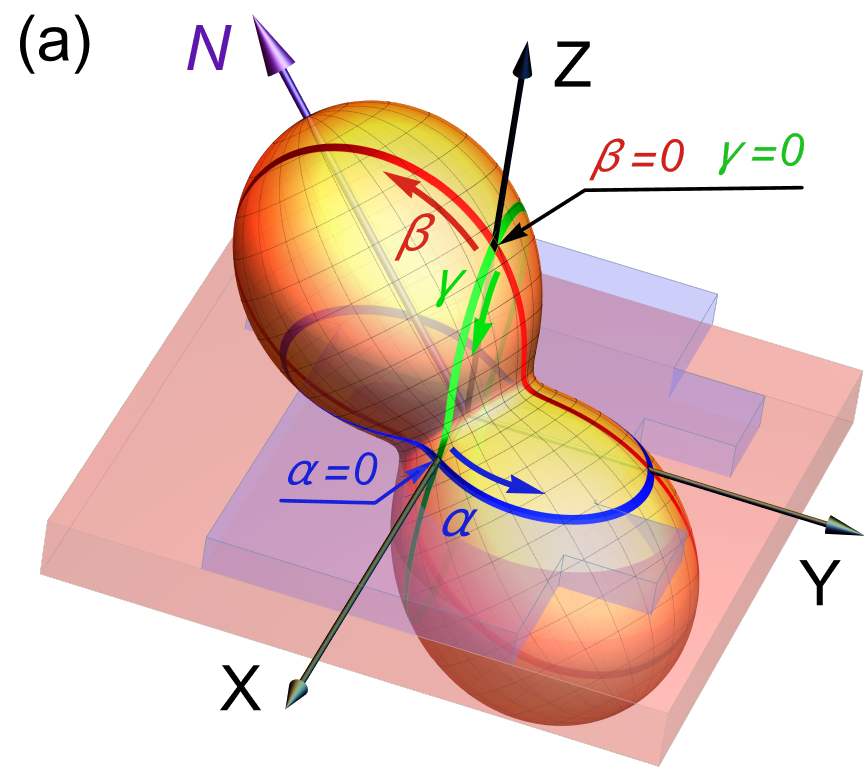}
\end{minipage}\hfill
\begin{minipage}{0.49\columnwidth}
    \centering
    \vspace{0pt}%
    \includegraphics[width = \textwidth]{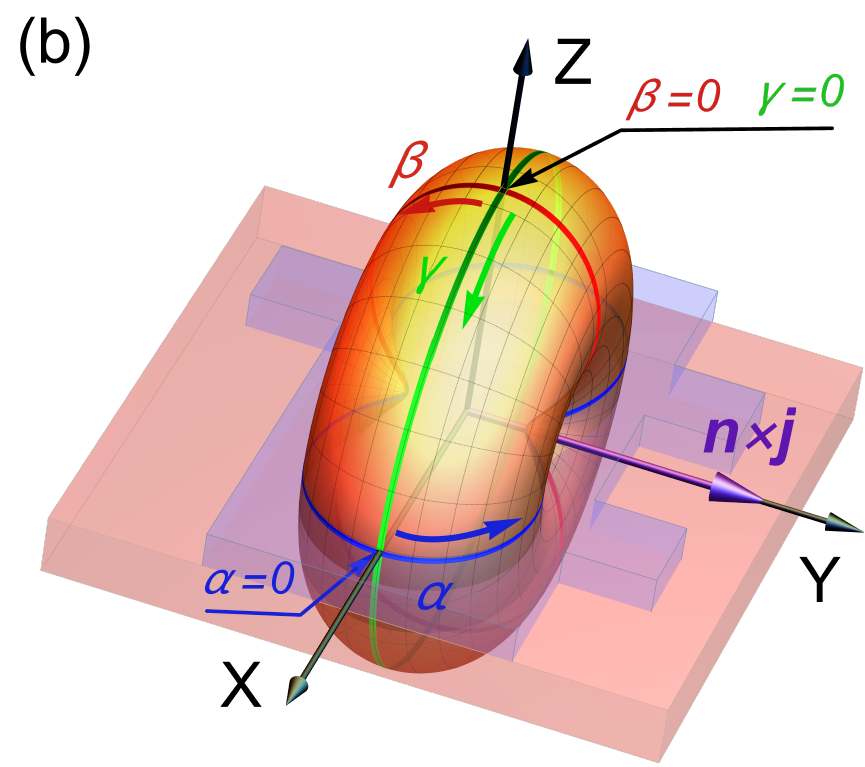}
\end{minipage}\\
\begin{minipage}{0.9\columnwidth}
    \centering
\includegraphics[width = 0.45\textwidth]{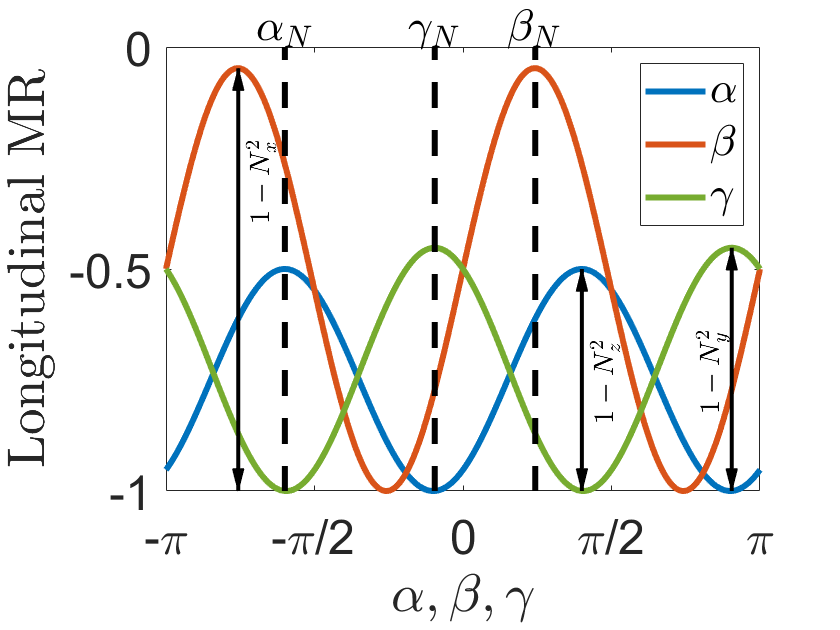}
    \includegraphics[width = 0.45\textwidth]{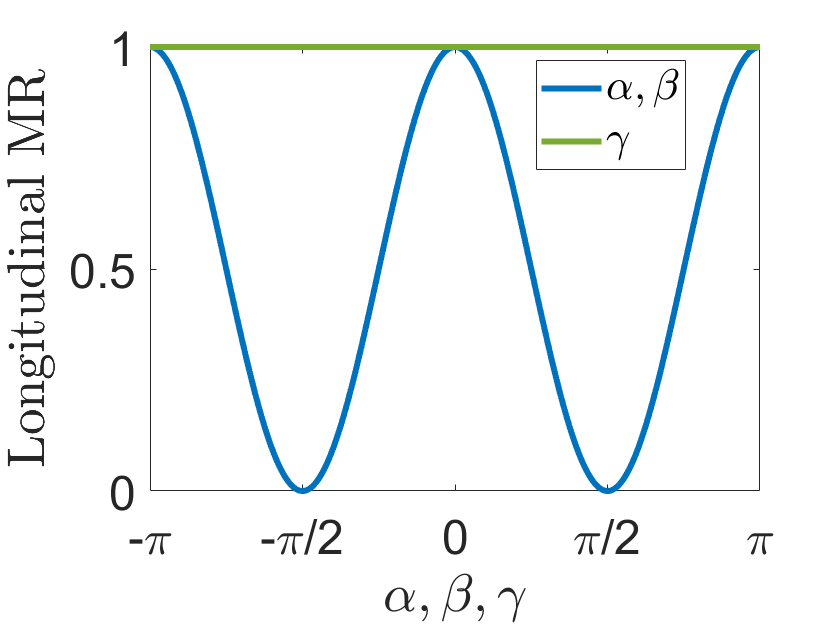}
    \includegraphics[width = 0.45\textwidth]{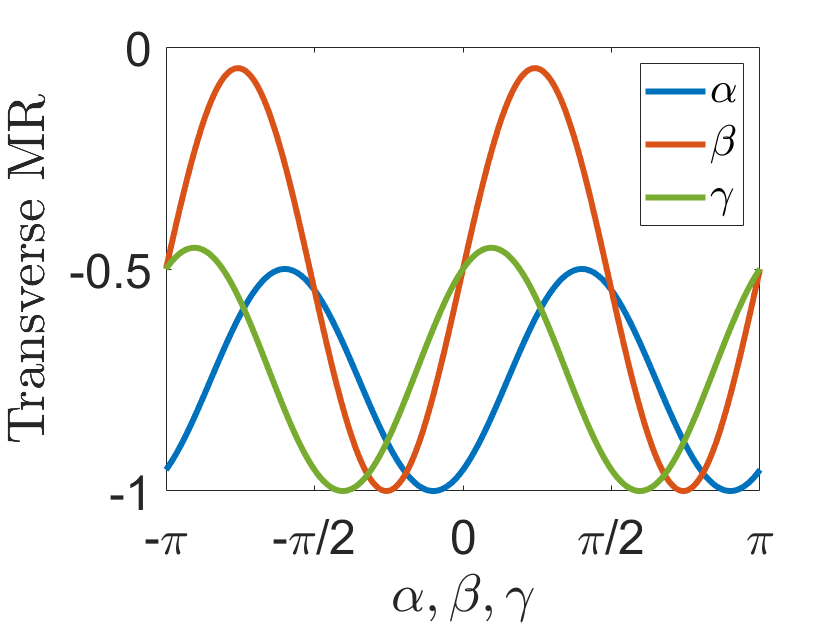}
    \includegraphics[width = 0.45\textwidth]{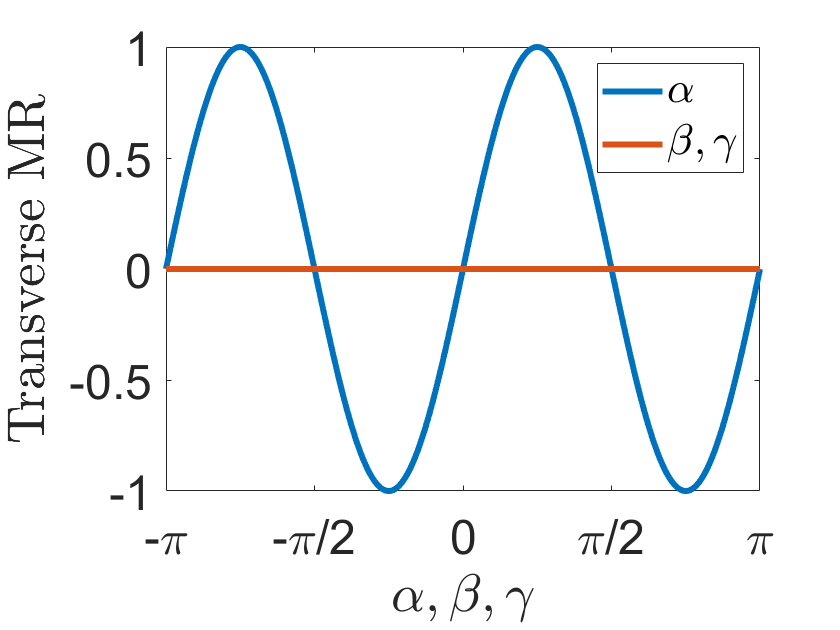}
\end{minipage}
\end{minipage}
    \caption{\label{fig:AngleDependence}
    Angular dependence of the longitudinal ($\rho_L$) and transverse ($\rho_T$)
    magnetoresistance for spin-splitter magnetoresistance (SSMR) in a collinear
    $d$-wave altermagnet
    (left column) and conventional spin Hall magnetoresistance (SMR, right column).
    \textbf{Top row:}
    Spherical plots of the longitudinal magnetoresistance, where the radial distance
    is proportional to $\rho_L$ (minus a constant)
    and the direction corresponds to the magnetization unit vector $\bm{m}$ of the FI.
    The opposite sign of the leading angular correction produces a peanut-shaped
    surface for SSMR and a doughnut-shaped surface for SMR.
    Its high-symmetry axis
    is set by the Néel vector $\bm{N}$ for SSMR
    and by the SHE vector $\bm{n}\times\bm{j}$ for SMR.
    \textbf{Middle row:}
    Line cuts of the longitudinal magnetoresistance correction
    through the standard $\alpha$, $\beta$, and $\gamma$
    rotation planes defined with respect to the Hall bar~\cite{nakayama2013spin}.
    Here and in the bottom row we subtract the angle-independent offset
    (the constant $\Delta\rho_0$) to isolate the angular variation;
    in both rows the longitudinal and transverse variations are each normalized
    to their maximum value.
    For SSMR this variation is $\Delta\rho_1\left[1-(\vec{m}\cdot\vec{N})^2\right]$,
    with $\Delta\rho_1<0$ [Eq.~(\ref{eq:Longitudinalrho1iso})],
    whereas for SMR it is $\Delta\rho_1^{\text{SMR}}\left(1-m_Y^2\right)$,
    with $\Delta\rho_1^{\text{SMR}}>0$.
    For SSMR, the angles $\alpha_N$, $\beta_N$, and $\gamma_N$
    mark the positions of the maxima used to extract $\bm{N}$.
    \textbf{Bottom row:}
    Corresponding line cuts of the transverse correction $\Delta\rho_T$.
    For SSMR it has the same angular dependence as $\Delta\rho_L$ up to a prefactor,
    whereas for SMR the two are qualitatively different,
    with the transverse variation given by $\Delta\rho_1^{\text{SMR}}\,m_X m_Y$.}
\end{figure}
The most distinctive experimental signature that separates SSMR from conventional
SMR appears when both the longitudinal and transverse magnetoresistances are measured. This difference is illustrated in the bottom panels of Fig.~\ref{fig:AngleDependence}.
For SSMR the transverse resistivity $\rho_T$ remains strictly proportional
to the longitudinal correction $\Delta\rho_L$ [Eq.~(\ref{eq:rhoLrhoT})].
In SMR the two responses are also related, but through a more intricate angular dependence~\cite{chen2013theory}.
For example, in the $\alpha$-plane the transverse signal is shifted by
$\pi/4$ relative to the longitudinal one for SMR,
while no such shift occurs for SSMR.
A complete ADMR characterization, including measurements of both $\rho_L$ and $\rho_T$ in the $\alpha$, $\beta$, and $\gamma$ rotation planes, therefore provides the clearest means of distinguishing SSMR from SMR.

A further distinction between SSMR and SMR can be obtained
by comparing two Hall bars on the same single-crystal domain but rotated by 90$^\circ$
with respect to each other.
In SMR the longitudinal resistance is invariant under this rotation and only the sign of the transverse resistance flips.
In SSMR the longitudinal resistances generally differ (being sensitive to different tensor components), while the transverse resistances remain equal.
This provides an additional experimental signature of the spin-splitter mechanism.

In a high-quality altermagnet with sizable spin splitting, spin relaxation is strongly anisotropic: spins oriented perpendicular to the Néel vector relax much faster than those aligned with it. To illustrate the consequences of this anisotropy, we consider the weakly diffusive limit $l_{s\perp}\ll l_{s\parallel}$ of the spin-relaxation tensor. Setting $l_{s\perp}\approx 0$ and neglecting spin-flip scattering at the interface ($G_s=0$), the longitudinal magnetoresistance simplifies to
\begin{align}
\Delta\rho_L = \frac{2T_{xy}^2l_{s\parallel}}{\sigma L_y}\frac{\sigma + G_r l_{s\parallel}\bigl[1-(\vec{m}\cdot\vec{N})^2\bigr]\coth\frac{L_y}{2l_{s\parallel}}}{\sigma+2G_r l_{s\parallel}\bigl[1-(\vec{m}\cdot\vec{N})^2\bigr]\coth\frac{L_y}{l_{s\parallel}}}\;.
\label{eq:Largetausperp}
\end{align}
\begin{figure}[!t]
    \centering
    \includegraphics[width=4.2cm]{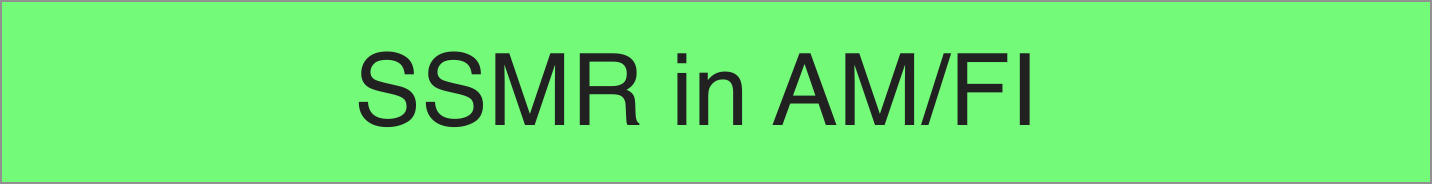}
    \includegraphics[width=4.2cm]{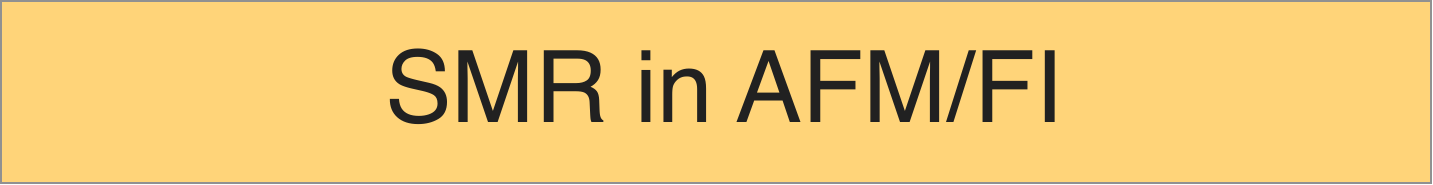}\\
    \vspace{8pt}
    \includegraphics[width=4.2cm]{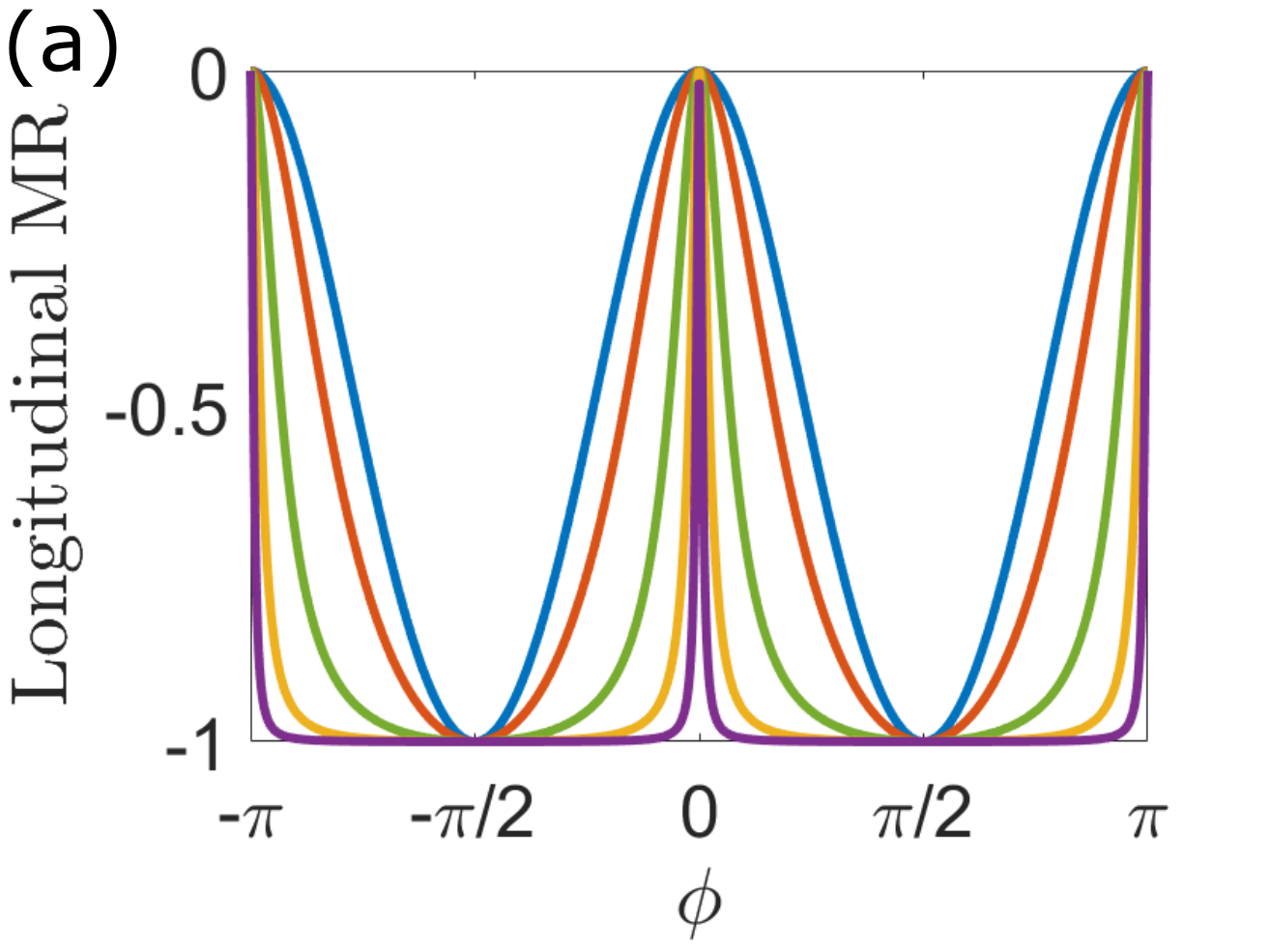}
    \includegraphics[width=4.2cm]{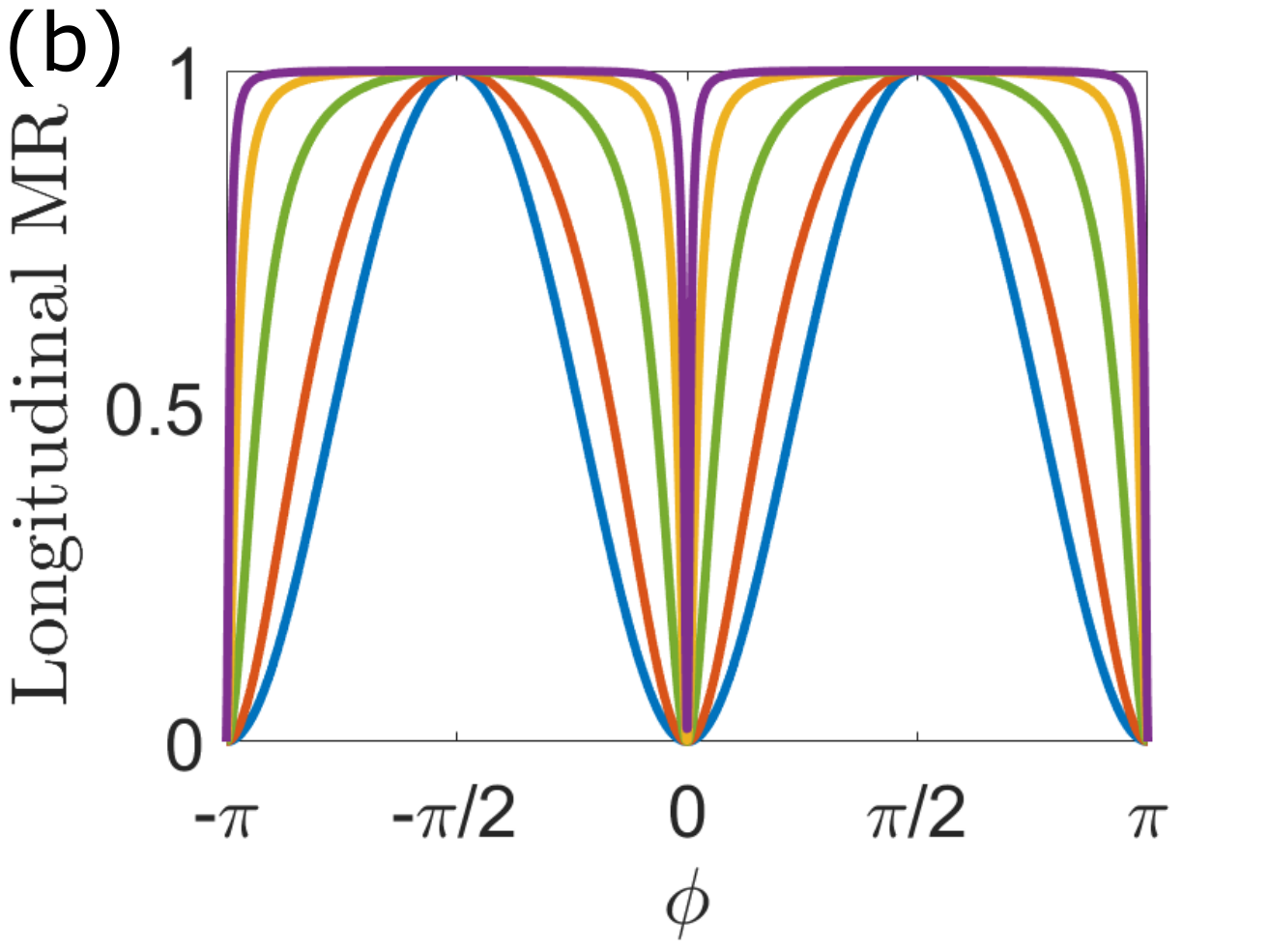}\\
    \vspace{6pt}
    \includegraphics[width=4.2cm]{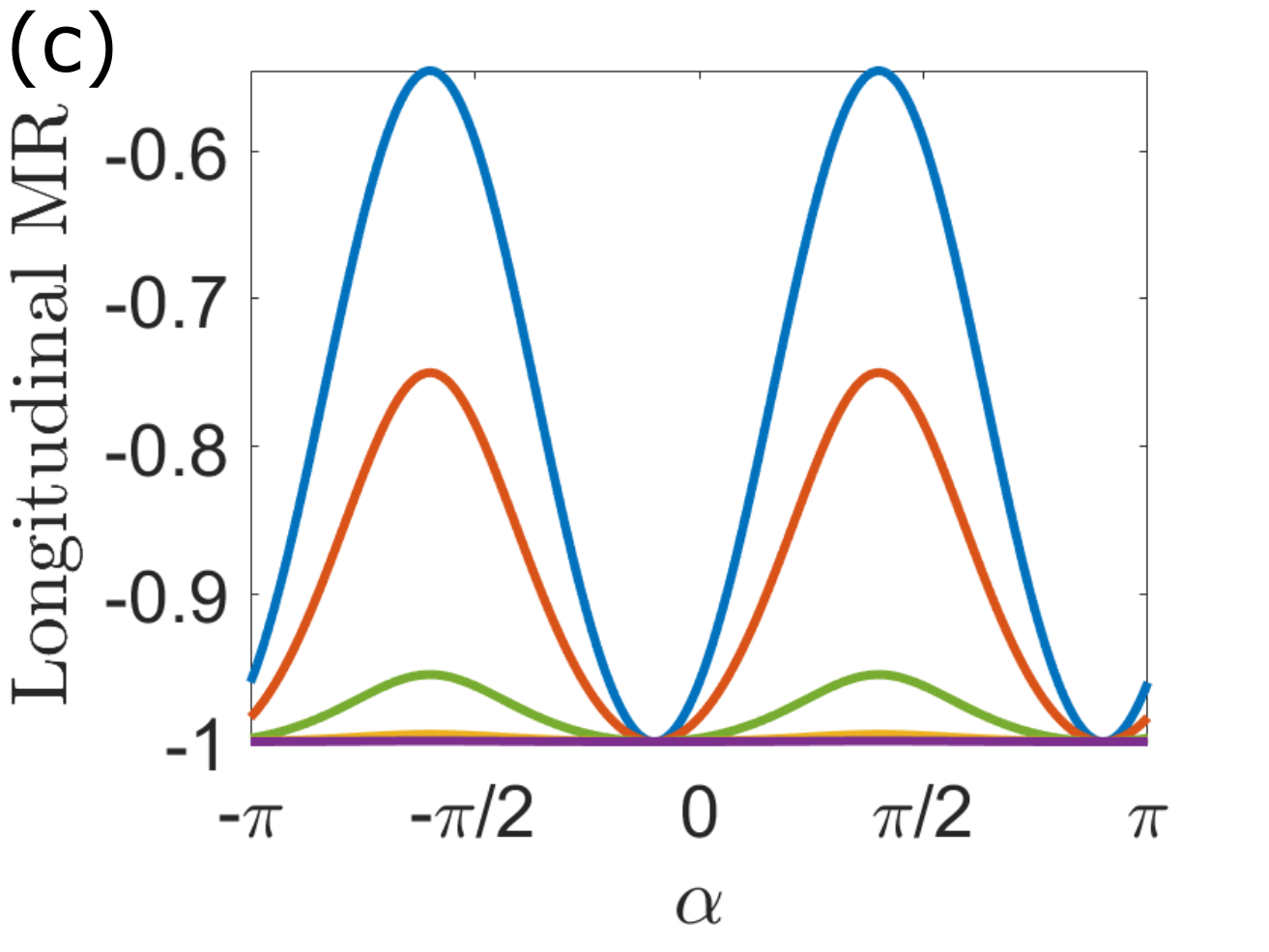}
    \includegraphics[width=4.2cm]{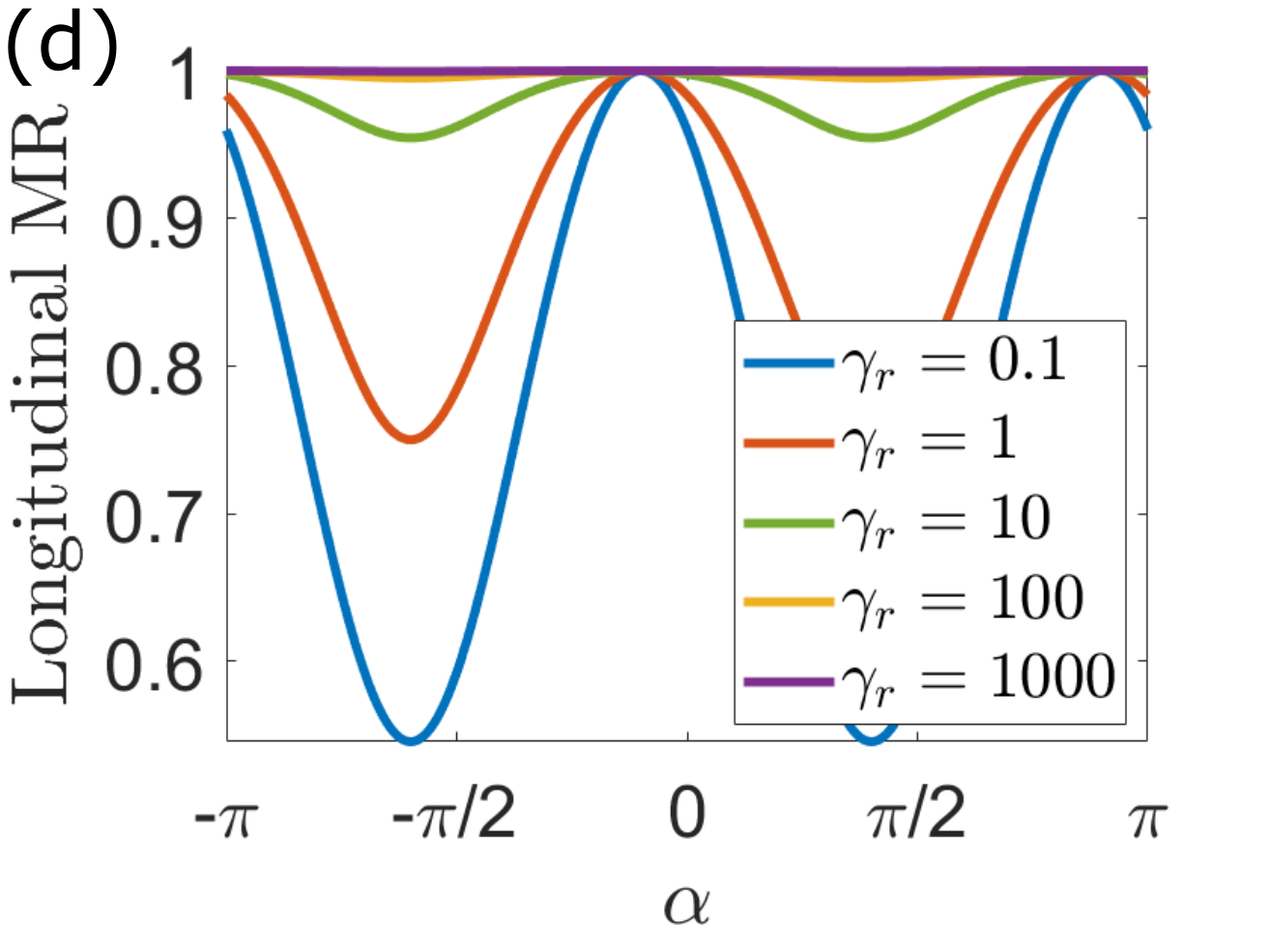}
    \caption{Anharmonic ADMR for SSMR in AM/FI (left) and SMR in AFM/FI (right).
    All panels show the longitudinal MR variation, defined and normalized
    as in Fig.~\ref{fig:AngleDependence}.
    \textbf{(a)} Dependence of the resistivity on the angle $\phi$ between the FI magnetization
    and the N\'eel vector of an altermagnet with strong relaxation of spins perpendicular
    to the N\'eel vector, at $G_s = 0$.
    For large $\gamma_r = G_rl_{s\parallel}/\sigma$
    the signal deviates strongly from the usual harmonic form,
    showing narrow peaks rising from broad plateaux.
    We set $L_y\gg l_{s\parallel}$, so that $\coth(L_y/l_{s\parallel})\approx 1$.
    \textbf{(b)} Same as (a), but for a metallic antiferromagnet with spin-orbit coupling,
    where the line shape is inverted: the curve instead shows narrow dips within broad plateaux.
    \textbf{(c,d)} Longitudinal MR as a function of $\alpha$
    for the N\'eel vector orientation as in Fig.~\ref{fig:AngleDependence}.
    When the magnetic-field sweep does not bring $\bm{m}$ sufficiently close to the N\'eel vector,
    the variation stays close to harmonic and becomes suppressed with increasing $\gamma_r$.}
    \label{fig:anharmonicity}
\end{figure}
While the overall angular dependence remains qualitatively similar to the isotropic case, the magnetoresistance is no longer purely harmonic: the denominator now depends explicitly on $(\vec{m}\cdot\vec{N})^2$. The strength of these higher-harmonic corrections is set by the dimensionless parameter $(G_r l_{s\parallel}/\sigma)\coth(L_y/l_{s\parallel})$. For a weakly coupled interface (small $G_r$), the anharmonicity is negligible. In the opposite, strong-coupling limit ($G_r l_{s\parallel}/\sigma\gg 1$), the resistivity develops narrow peaks whenever $\vec{m}$ is parallel or antiparallel to $\vec{N}$. Figure~\ref{fig:anharmonicity}(a) shows how these peaks sharpen with increasing $G_r$.
Interestingly, the same anisotropic relaxation produces the opposite effect in conventional SMR on antiferromagnets, where the peaks broaden and the dips sharpen [Fig.~\ref{fig:anharmonicity}(b)].

As the SSMR peaks narrow, they become increasingly difficult to resolve, because the ADMR amplitude depends sensitively on how closely $\vec{m}$ approaches $\vec{N}$. Most of the variation is concentrated within an angular window around the Néel vector, set by $(G_r l_{s\parallel}/\sigma)\coth(L_y/l_{s\parallel})\sin^2\phi\lesssim 1$. 
Detecting this narrowing therefore
requires a magnetic-field sweep that brings $\vec{m}$ sufficiently close to the Néel-vector orientation;
in other sweep planes the variation is suppressed and shows no clear narrowing, as illustrated in Fig.~\ref{fig:anharmonicity}(c)
for the $\alpha$-plane and the Néel orientation of the top-left panel of Fig.~\ref{fig:AngleDependence}.

Note that only the real part $G_r$ of the spin-mixing conductance enters Eq.~(\ref{eq:Largetausperp});
the imaginary part $G_i$ drops out entirely.
This underscores the importance of engineering interfaces with a moderately large $G_r$
if SSMR is to be observed in the weakly diffusive regime,
where the spin-relaxation tensor $\Gamma_{ab}$ is strongly anisotropic.

To distinguish altermagnets from conventional metallic
antiferromagnets---both being metallic---it
is instructive to compare the altermagnetic SSMR with the SMR arising from
the SHE in collinear antiferromagnets coupled to FIs.
We focus on the regime of large longitudinal spin fluctuations,
where $l_{s\perp}\ll l_{s\parallel}$.
\begin{table*}[!t]%
\centering %
\caption{Comparison of distinguishing features of spin Hall magnetoresistance (SMR) in nonmagnetic metals (NM) and metallic antiferromagnets (AFM) versus spin-splitter magnetoresistance (SSMR) in altermagnets (AM). The three effects differ in sign, symmetry axis, presence of anharmonic corrections, and magnitude scaling.\label{tab:NMAFMAM}}%
\begin{tabular*}{\textwidth}{@{\extracolsep\fill}lcccl@{\extracolsep\fill}}
\toprule
\textbf{Effect} & \textbf{Sign}  & \textbf{Axis}  & \textbf{Anharmonicity}  & \textbf{Magnitude} \\
\midrule
SMR in NM/FI & $+$ & $\vec{n}\times \vec{j}$ & \xmark & $\theta^2$   \\
SMR in AFM/FI & $+$ & $\vec{N}$ & \cmark & $\theta^2 (\vec{N}\cdot(\vec{n}\times \vec{j}))^2$   \\
SSMR in AM/FI & $-$ & $\vec{N}$ & \cmark & $T_{xy}^2$   \\
\bottomrule
\end{tabular*}
\end{table*}
As shown in Appendix~\ref{sec:SMRAFM}, in this limit the SMR in antiferromagnets
shares several qualitative features with SSMR:
the resistance depends on the angle between the magnetization of the FI
and the Néel vector, the longitudinal and transverse resistivities exhibit
the same angular dependence,
and anisotropic relaxation produces anharmonic corrections.
At the same time, it retains the chiral sign characteristic of
conventional SMR in nonmagnetic metals, which is opposite to that of SSMR.
Consequently, anisotropic relaxation modifies the line shape in the opposite way:
the peaks broaden and the dips sharpen, in contrast to the narrowing of peaks observed in SSMR.
In addition, the magnitude of this antiferromagnetic SMR depends on
the relative orientation between the Néel vector and $\bm{n}\times\bm{j}$.
Away from this limit, when $l_{s\perp}\approx l_{s\parallel}$,
the role of the Néel vector is strongly diminished
and the response crosses over to conventional SMR.
These contrasting signatures allow the SMR in metallic antiferromagnets to be readily distinguished both from conventional SMR and from SSMR in altermagnets.
We summarize the main differences between these three effects in Table~\ref{tab:NMAFMAM}.

\section{Conclusions}

In summary, we have developed a theory of angular-dependent magnetoresistance in metallic $d$-wave altermagnets coupled to ferromagnetic insulators, and we have identified the spin-splitter magnetoresistance as a smoking-gun signature of collinear altermagnetism. In contrast to conventional SMR, the SSMR displays three robust and experimentally accessible distinctions: (i) it depends exclusively on the relative orientation of the ferromagnetic magnetization $\bm{m}$ and the altermagnetic N\'eel vector $\bm{N}$; (ii) its longitudinal ADMR response carries the opposite sign; and (iii) its transverse resistivity is directly proportional to the longitudinal correction. As we have shown, all three originate from the non-chiral, band-structure-driven character of the spin-splitter effect, and they persist across both the isotropic and the strongly anisotropic spin-relaxation regimes.

Beyond identifying the effect, our results turn the full angular dependence of the longitudinal and transverse signals into a practical tool: measuring the ADMR in the standard $\alpha$, $\beta$, and $\gamma$ rotation planes allows the orientation of $\bm{N}$ to be extracted (up to its sign) and altermagnets to be distinguished unambiguously from both nonmagnetic metals and metallic antiferromagnets. In the latter case the SMR retains the chiral sign of conventional SMR while acquiring anharmonic corrections reminiscent of the SSMR, so that a joint comparison of sign, symmetry axis, and line shape, summarized in Table~\ref{tab:NMAFMAM}, separates the three cases.

More broadly, these results establish the SSMR as a distinct class of magnetoresistance, separate from spin-orbit-driven effects such as SMR and HMR, and they offer concrete design guidelines for future experiments. We anticipate that a complete ADMR characterization of high-quality altermagnet/ferromagnetic-insulator bilayers---ideally in materials whose bulk altermagnetic order is firmly established---will provide one of the cleanest transport routes both to confirming altermagnetism and to mapping the N\'eel vector in novel materials.

\bmsection*{Acknowledgments}
We thank Tero Heikkilä for useful discussions.
T. K. acknowledges support by the Research Council of Finland through DYNCOR, Project Number 354735 and through the Finnish Quantum Flagship, Project Number 359240.  Moreover, his work is part of the Finnish Centre of Excellence in Quantum Materials (QMAT). F.~S.~B. thanks financial support from the Spanish MCIN/AEI/10.13039/501100011033
through the grant PID2023-148225NB-C31, and the
European Union’s Horizon Europe research and innovation program under grant agreement No. 101130224 (JOSEPHINE).
V.N.G. acknowledges funding from the Spanish MCIN/AEI/10.13039/501100011033 through grant PID2024-160189NA-I00,
co-funded by the European Regional Development Fund (FEDER, EU),
as well as from the Basque Government through grants PIBA$\_$2025$\_$1$\_$0018,
IT-1591-22, and IKUR/RESONANT.

\bmsection*{Conflict of interest}
The authors declare no conflict of interest.

\bibliographystyle{unsrtnat}
\bibliography{sources}

\clearpage
\appendix
\begin{strip}
\bmsection{Derivation of the SSMR for arbitrary spin-relaxation anisotropy}\label{sec:SSMRderivation}
In this Appendix we provide the details of the calculation of the spin-splitter magnetoresistance, SSMR. Specifically, starting from the charge conservation and spin diffusion equations, Eqs. (\ref{eq:ChargeCurrentconservatio}) and (\ref{eq:spindiffusionequation}) we derive Eq. (\ref{eq:Deltarhogeneral}) for the longitudinal resistance correction and Eq. (\ref{eq:rhoLrhoT}) for the relation between the longitudinal and transverse resistance in the main text. Moreover, we show how the limits of isotropic relaxation, Eqs. (\ref{eq:Longitudinal})-(\ref{eq:Longitudinalrho1iso}) in the main text, and highly anisotropic relaxation, Eq. (\ref{eq:Largetausperp}) in the main text, follow directly from our general expression.

We consider a geometry in which an altermagnet carries a current in its infinite $x$ direction, has a boundary with air at $y = \frac{L_y}{2}$ and a boundary with an FI at $y = -\frac{L_y}{2}$. By translational invariance  while $\mu$ satisfies \begin{align}\mu = -Ex+\delta\mu(y)\;,\end{align} where $E$ is a scalar that corresponds to the electric field. Moreover, because of spin-relaxation, translational invariance requires that the spin chemical potential $\vec{\mu}^s$  depends only on $y$.

Exploiting this translational invariance, Eqs. (\ref{eq:ChargeCurrentconservatio}) and (\ref{eq:spindiffusionequation}) reduce to effectively one-dimensional equations. At the boundary at $y = \frac{L_y}{2}$ both charge current and spin current are vanishing. At the boundary at $y = -\frac{L_y}{2}$ the charge current is vanishing, since we consider a ferromagnetic insulator, but the spin current satisfies Eq. (\ref{eq:FIBC}) in the main text.

For the optimal orientation for the SSMR, in which $T_{yy} = 0$, the equations for $\delta\mu$ and $\mu^s$ are decoupled. For $\delta\mu$ we obtain
\begin{align}
    \partial_{yy}\delta\mu &= 0\;,\\
    \partial_y\delta\mu &= 0\;, &(y = \pm \frac{L_y}{2})\;.
\end{align}
Thus, $\delta\mu$ is a constant, which may be set to zero by choice of the origin for the x-coordinate.
For $\vec{\mu}^s$ we have the following boundary value problem
\begin{align}
    D\partial_{yy}\vec{\mu}^s  &= \frac{1}{\tau_{s\parallel}}(\vec{\mu}_s\cdot\vec{N})\vec{N}+\frac{1}{\tau_{s\perp}}\vec{N}\times(\vec{\mu}_s\times\vec{N})\;,\\
    D\partial_{y}\vec{\mu}^s(y = \frac{L_{y}}{2}) &= 2T_{xy} D E \vec{N}\;,\\
    D\partial_{y}\vec{\mu}^s(y = -\frac{L_{y}}{2}) &= 2T_{xy} D E \vec{N} +D\frac{2|G_s|}{\sigma}\vec{\mu}^s- D\frac{2G_{r}}{\sigma}\vec{m}\times(\vec{m}\times\vec{\mu}^s)-D\frac{2G_{i}}{\sigma}\vec{m}\times\vec{\mu}^s.
\end{align}

This is a set of three one-dimensional second order diffusion equations with constant coefficients.
The bulk equation and the boundary condition at $y = \frac{L_y}{2}$ dictate that the solution can be written as
\begin{align}
    \vec{\mu}^s &= (2T_{xy}El_{s\parallel} \frac{\sinh{\frac{y}{l_{s\parallel}}}}{\cosh{\frac{L_y}{2l_{s\parallel}}}}+ C_1\frac{\cosh{\frac{y-\frac{L_y}{2}}{l_{s\parallel}}}}{\sinh{\frac{L_y}{l_{s\parallel}}}})\vec{N}+\vec{C}_2\frac{\cosh{\frac{y-\frac{L_y}{2}}{l_{s\perp}}}}{\sinh{\frac{L_y}{l_{s\perp}}}}\;,\label{eq:muscs}
\end{align}
where $\vec{C}_2\cdot\vec{N} = 0$.

From this solution, the charge current in the x-direction can be written as
\begin{align}
    J_x = L_y\sigma E -\frac{\sigma}{2}T_{xy}(\vec{N}\cdot\vec{\mu}^s(y = \frac{L_y}{2})-\vec{N}\cdot\vec{\mu}^s(y = \frac{L_y}{2})) = L_y\sigma E-2\sigma T_{xy}^2E l_{s\parallel}\tanh{\frac{L_y}{2l_{s\parallel}}}+\frac{1}{2}\sigma T_{xy}C_1\tanh{\frac{L_y}{2l_{s\parallel}}}\;.\label{eq:CurrentintermsofC1}
\end{align}
Thus, the SSMR is determined by the coefficient $C_1$ in Eq. (\ref{eq:muscs}).

To determine $C_1$ and $\vec{C}_2$, we consider the boundary condition at $y = -\frac{L_y}{2}$. It is convenient to make a change of basis to the following basis. We write $\vec{\mu}^s = \mu^s_m \vec{m} + \mu^s_{\perp1}\vec{a}+\mu^s_{\perp2}\vec{b}$, where $\vec{a},\vec{b}$ are unit vectors perpendicular to $\vec{m}$ such that $\vec{N} = (\vec{m}\cdot\vec{N})\vec{m}+(\vec{a}\cdot\vec{N})\vec{a}$ and $\vec{b} = \vec{m}\times\vec{a}$. Next we define the complex quantity $\mu^s_{\perp} = \mu^s_{\perp1}+i\mu^s_{\perp 2}$.

In this basis, the general solution can be written as
\begin{align}
    \mu^s_m &= (2T_{xy}El_{s\parallel} \frac{\sinh{\frac{y}{l_{s\parallel}}}}{\cosh{\frac{L_y}{2l_{s\parallel}}}}+ C_1\frac{\cosh{\frac{y-\frac{L_y}{2}}{l_{s\parallel}}}}{\sinh{\frac{L_y}{l_{s\parallel}}}})\cos\phi + \text{Re}(C_3)\frac{\cosh{\frac{y-\frac{L_y}{2}}{l_{s\perp}}}}{\sinh{\frac{L_y}{l_{s\perp}}}}\sin\phi\;,\label{eq:musm}\\
    \mu^s_\perp &= (2T_{xy}El_{s\parallel} \frac{\sinh{\frac{y}{l_{s\parallel}}}}{\cosh{\frac{L_y}{2l_{s\parallel}}}}+ C_1\frac{\cosh{\frac{y-\frac{L_y}{2}}{l_{s\parallel}}}}{\sinh{\frac{L_y}{l_{s\parallel}}}})\sin\phi - (\text{Re} C_3\cos\phi + i\text{Im} C_3) \frac{\cosh{\frac{y-\frac{L_y}{2}}{l_{s\perp}}}}{\sinh{\frac{L_y}{l_{s\perp}}}}\label{eq:musp}\;,
\end{align}
where $\phi$ is the angle between $\vec{m}$ and $\vec{N}$, $C_3$ is a complex number such that $\text{Re}(C_3) = \sqrt{(\vec{m}\cdot\vec{C}_2)^2 + (\vec{a}\cdot\vec{C}_2)^2}$ and $\text{Im}(C_3) = \vec{b}\cdot\vec{C}_2$, while $C_1$ is the same real number as in Eq. (\ref{eq:muscs}).

The boundary condition at $y = -\frac{L_y}{2}$ in this basis reads
\begin{align}
    \sigma\partial_y \mu^s_m&= 2T_{xy}E \cos\phi +2|G_s|\mu^s_m\;,\label{eq:BCmusm}\\
    \sigma\partial_y \mu^s_\perp&=2T_{xy}E \sin\phi + 2(|G_s|+G_r-iG_i)\mu^s_\perp\;.\label{eq:BCmusp}
\end{align}
Now, for short-hand notation we define $G_{\parallel,\perp} = \frac{\sigma}{l_{s\parallel,\perp}}$ and $\mu^s_0 = 2T_{xy}El_{s\parallel}\tanh{\frac{L_y}{2l_{s\parallel}}}$.
Substituting Eqs. (\ref{eq:musm},\ref{eq:musp}) into Eqs. (\ref{eq:BCmusm},\ref{eq:BCmusp}) we find
\begin{align}
    G_{\parallel}C_1 \cos\phi + G_{\perp}\text{Re}(C_3)\sin\phi = 2|G_s|\Big((\mu^s_0-C_1\coth{\frac{L_y}{2l_{s\parallel}}})\cos\phi - \text{Re}(C_3)\coth{\frac{L_y}{2l_{s\perp}}})\sin\phi\Big)\;,\label{eq:Firsteq}\\
    G_{\parallel}C_1\sin\phi - G_{\perp}C_3 \cos\phi = 2(|G_s| + G_r-iG_i)\Big((\mu_0^s-C_1\coth{\frac{L_y}{2l_{s\parallel}}})\sin\phi+(\text{Re}(C_3)\cos\phi + i\text{Im}C_3)\coth{\frac{L_y}{2l_{s\perp}}})\Big)\;.
\end{align}

Notably, the first equation, only involves $C_1$ and $\text{Re}(C_3)$, while the second equation may be rewritten as
\begin{align}\label{eq:Secondeq}
    \text{Re}(C_3)\cos\phi + i\text{Im}(C_3)-(\frac{G_{\parallel}+2(|G_s|+G_r-G_i)\coth{\frac{L_y}{l_{s\parallel}}}}{G_{\perp}+2(|G_s|+G_r-G_i)\coth{\frac{L_y}{l_{s\perp}}}})C_1 = -2\mu_0^s\sin\phi \frac{|G_s|+ G_{r}-iG_i}{G_{\perp}+2(|G_s|+G_r-G_i)\coth{\frac{L_y}{l_{s\perp}}}}\;.
\end{align}
By taking the real part of this equation one obtains an equation that involves $C_1$ and $\text{Re}(C_3)$. Thus, $C_1$ and $\text{Re}(C_3)$ can be determined by the system of Eq. (\ref{eq:Firsteq}) and the real part of (\ref{eq:Secondeq}). The imaginary part of Eq. (\ref{eq:Secondeq}) can then in principle be used to determine $\text{Im}(C_3)$. However, since the SSMR only depends on $C_1$, we do not need to do so.

With this, we have reduced the problem to a set of two linear equations:
\begin{align}
    (G_{\parallel}+ 2|G_s|\coth{\frac{L_y}{2l_{s\parallel}}})\cos\phi C_1 + (G_{\perp}+ 2|G_s|\coth{\frac{L_y}{2l_{s\perp}}})\sin\phi\text{Re}(C_3) = 2 |G_s|\mu_0^s \cos\phi\;,
\end{align}

\begin{align}
    \text{Re}(C_3)\cos\phi-\text{Re}(\frac{G_{\parallel}+2(|G_s|+G_r-G_i)\coth{\frac{L_y}{l_{s\parallel}}}}{G_{\perp}+2(|G_s|+G_r-G_i)\coth{\frac{L_y}{l_{s\perp}}}})C_1 = -2\mu_0^s\sin\phi\text{Re}\frac{|G_s|+ G_{r}-iG_i}{G_{\perp}+2(|G_s|+G_r-G_i)\coth{\frac{L_y}{l_{s\perp}}}}\;.
\end{align}

Introducing $\mathcal{R}(G) = \text{Re}\Big(\frac{2G}{G_{\perp}+2G\coth{\frac{L_y}{l_{s\perp}}}}\Big) = \text{Re}\Big(\frac{2Gl_{s\perp}}{\sigma+2Gl_{s\perp}\coth{\frac{L_y}{l_{s\perp}}}}\Big)$, this can be written as,

\begin{align}
    \text{Re}(C_3)\sin\phi + (\frac{G_{\parallel}}{G_{\perp}}+\mathcal{R}(|G_s|)(\coth{\frac{L_y}{l_{s\parallel}}}-\frac{G_{\parallel}}{G_{\perp}}\coth{\frac{L_y}{l_{s\perp}}})C_1\cos\phi &= \mu_0^s\cos\phi\mathcal{R}(|G_s|)\;,\\
    \text{Re}(C_3)\cos\phi-(\frac{G_{\parallel}}{G_{\perp}}+
    \mathcal{R}(|G_s|+G_{\uparrow\downarrow})(\coth{\frac{L_y}{l_{s\parallel}}}-\frac{G_{\parallel}}{G_{\perp}}\coth{\frac{L_y}{l_{s\perp}}})C_1\sin\phi &= -\mu_0^s\mathcal{R}(|G_s|+G_{\uparrow\downarrow})\sin\phi\;,
\end{align}

Eliminating $\text{Re}(C_3)$, we find
\begin{align}
    C_1 & = \mu_0^s\frac{\mathcal{R}(|G_s|) + (\mathcal{R}(|G_s|+G_{\uparrow\downarrow})-\mathcal{R}(|G_s|))\sin^2\phi}{\frac{
    l_{s\perp}}{l_{s\parallel}}+\Big(\mathcal{R}(|G_s|) + (\mathcal{R}(|G_s|+G_{\uparrow\downarrow})-\mathcal{R}(|G_s|))\sin^2\phi\Big)(\coth{\frac{L_y}{l_{s\parallel}}}-\frac{
    l_{s\perp}}{l_{s\parallel}}\coth{\frac{L_y}{l_{s\perp}}})}\;.
\end{align}
Next, we may exploit

\begin{align}
    &\mathcal{R}(|G_s|+G_{\uparrow\downarrow})-\mathcal{R}(|G_s|) = \text{Re}\frac{-2|G_s|(\sigma + 2(|G_s|+G_{\uparrow\downarrow}) l_{s\perp}\coth{\frac{L_y}{l_{s\perp}}})+2(|G_s|+G_{\uparrow\downarrow})(\sigma + 2|G_s| l_{s\perp}\coth{\frac{L_y}{l_{s\perp}}})}{(\sigma + 2|G_s| l_{s\perp}\coth{\frac{L_y}{l_{s\perp}}})(\sigma + 2(|G_s|+G_{\uparrow\downarrow}) l_{s\perp}\coth{\frac{L_y}{l_{s\perp}}})} \nonumber\\&= \text{Re}\frac{1}{\sigma + 2|G_s| l_{s\perp}\coth{\frac{L_y}{l_{s\perp}}}}\frac{G_{\uparrow\downarrow}l_{s\perp}}{1+2\frac{|G_s|+G_{\uparrow\downarrow}}{\sigma}l_{s\perp}\coth{\frac{L_y}{l_{s\perp}}}} =  \frac{2\bar{G}l_{s\perp}}{\sigma + 2G_s l_{s\perp}\coth{\frac{L_y}{l_{s\perp}}}}\;,
\end{align}
where we define the effective conductance
\begin{align}
    \bar{G} = \text{Re}\frac{G_{\uparrow\downarrow}}{1+2\frac{|G_s|+G_{\uparrow\downarrow}}{\sigma}l_{s\perp}\coth{\frac{L_y}{l_{s\perp}}}}\;.
\end{align}
We may now rewrite $C_1$ as
\begin{align}\label{eq:C1}
    C_1 &= \mu_0^s\frac{2(|G_s|+\bar{G}\sin^2\phi) l_{s\perp}}{\frac{l_{s\perp}}{l_{s\parallel}}(\sigma+2G_sl_{s\perp}\coth\frac{L_y}{l_{s\perp}})+2(|G_s|+\bar{G}\sin^2\phi)(\coth{\frac{L_y}{l_{s\parallel}}}-\frac{
    l_{s\perp}}{l_{s\parallel}}\coth{\frac{L_y}{l_{s\perp}}})}\\
    &=\mu_0^s\frac{2\frac{l_{s\parallel}}{\sigma}(|G_s|+\bar{G}\sin^2\phi )}{1+\frac{2|G_s|l_{s\perp}}{\sigma}\coth\frac{L_y}{l_{s\perp}}+2(|G_s|+\bar{G}\sin^2\phi)(\frac{l_{s\parallel}}{\sigma}\coth{\frac{L_y}{l_{s\parallel}}}-\frac{
    l_{s\perp}}{\sigma}\coth{\frac{L_y}{l_{s\perp}}})}\\
    &=\mu_0^s\frac{\frac{l_{s\parallel}}{\sigma}2(|G_s|+\bar{G}\sin^2\phi )}{1+\frac{2|G_s|l_{s\parallel}}{\sigma}\coth\frac{L_y}{l_{s\parallel}}+2(\frac{l_{s\parallel}}{\sigma}\coth{\frac{L_y}{l_{s\parallel}}}-\frac{
    l_{s\perp}}{\sigma}\coth{\frac{L_y}{l_{s\perp}}})\bar{G}\sin^2\phi}\;.
\end{align}
As discussed before, $C_1$ determines the SSMR strength, through Eq. (\ref{eq:CurrentintermsofC1}). Indeed, Eq. (\ref{eq:CurrentintermsofC1}) implies
\begin{align}\label{eq:deltaJ1}
    \delta J = J-\sigma L_y E = -T_{xy}\sigma (\mu_0^s-\frac{1}{2}\tanh{\frac{L_y}{2l_{s\parallel}}}C_1)\;.
\end{align}
Substituting Eq. (\ref{eq:C1}) into Eq. (\ref{eq:deltaJ1}) and exploiting that $2\coth{\frac{L_y}{l_{s\parallel}}}-\tanh{\frac{L_y}{2l_{s\parallel}}} = \coth{\frac{L_y}{2l_{s\parallel}}}$, we find
\begin{align}
    \delta J& = -T_{xy}\mu_0^s (1-\frac{1}{2}\mu_0^s\frac{2\frac{l_{s\parallel}}{\sigma}\tanh{\frac{L_y}{2l_{s\parallel}}}(|G_s|+\bar{G}\sin^2\phi) }{1+\frac{2|G_s|l_{s\parallel}}{\sigma}\coth\frac{L_y}{l_{s\parallel}}+2(\frac{l_{s\parallel}}{\sigma}\coth{\frac{L_y}{l_{s\parallel}}}-\frac{
    l_{s\perp}}{\sigma}\coth{\frac{L_y}{l_{s\perp}}})\bar{G}\sin^2\phi}\\
    &= -T_{xy}\sigma\mu_0^s\frac{1+\frac{|G_s|l_{s\parallel}}{\sigma}\coth\frac{L_y}{2l_{s\parallel}}+(\frac{l_{s\parallel}}{\sigma}\coth{\frac{L_y}{2l_{s\parallel}}}-2\frac{
    l_{s\perp}}{\sigma}\coth{\frac{L_y}{l_{s\perp}}})\bar{G}\sin^2\phi}{1+\frac{2|G_s|l_{s\parallel}}{\sigma}\coth\frac{L_y}{l_{s\parallel}}+2(\frac{l_{s\parallel}}{\sigma}\coth{\frac{L_y}{l_{s\parallel}}}-\frac{
    l_{s\perp}}{\sigma}\coth{\frac{L_y}{l_{s\perp}}})\bar{G}\sin^2\phi}\;,
\end{align}
so that, substituting $\mu^s_0 = 2T_{xy}El_{s\parallel}\tanh{\frac{L_y}{2l_{s\parallel}}}$ and using $\frac{\delta J}{J} = -\frac{\Delta \rho_L}{\rho}$, we find
\begin{align}\label{eq:DeltarhogeneralSI}
    \frac{\Delta \rho_L}{\rho_{L}} = 2T_{xy}^2\frac{l_{s_{\parallel}}}{L_y}\tanh{\frac{L_y}{2l_{s\parallel}}}\frac{1+\frac{|G_s|l_{s\parallel}}{2\sigma}\coth\frac{L_y}{2l_{s\parallel}}+(\frac{l_{s\parallel}}{2\sigma}\coth{\frac{L_y}{2l_{s\parallel}}}-\frac{
    l_{s\perp}}{\sigma}\coth{\frac{L_y}{l_{s\perp}}})\bar{G}\sin^2\phi}{1+\frac{|G_s|l_{s\parallel}}{\sigma}\coth\frac{L_y}{l_{s\parallel}}+(\frac{l_{s\parallel}}{\sigma}\coth{\frac{L_y}{l_{s\parallel}}}-\frac{
    l_{s\perp}}{\sigma}\coth{\frac{L_y}{l_{s\perp}}})\bar{G}\sin^2\phi}\;.
\end{align}
Multiplying both sides with $\rho_L = \sigma^{-1}$, we recover Eq. (\ref{eq:Deltarhogeneral}) in the main text.

Moreover, we have
\begin{align}
    J_z = \int_{-\frac{L_y}{2}}^{\frac{L_y}{2}}-\sigma (\partial_z \mu + T_{yz}\vec{N}\cdot\partial_y\vec{\mu}^s+ T_{xz}\vec{N}\cdot\partial_x \vec{\mu_s}) = \frac{T_{yz}}{T_{xy}}\delta J\;.
\end{align}
Therefore, we have
\begin{align}
    \rho_T = \frac{T_{yz}}{T_{xy}}\Delta \rho_{L}\;.
\end{align}
This corresponds to the second equality in Eq. (\ref{eq:rhoLrhoT}) in the main text.

We may now also analyse how the special cases, isotropic relaxation, Eqs. (\ref{eq:Longitudinal}-\ref{eq:Longitudinalrho1iso}) in the main text, and highly anisotropic relaxation, Eq. (\ref{eq:Largetausperp}) in the main text, follow directly from this expression.
First consider the limit of isotropic relaxation, i.e. $l_{s\parallel} = l_{s\perp} = l_s$. In that case $(\frac{l_{s\parallel}}{\sigma}\coth{\frac{L_y}{l_{s\parallel}}}-\frac{
    l_{s\perp}}{\sigma}\coth{\frac{L_y}{l_{s\perp}}})\rightarrow{}0$ and $(\frac{l_{s\parallel}}{2\sigma}\coth{\frac{L_y}{2l_{s\parallel}}}-\frac{
    l_{s\perp}}{\sigma}\coth{\frac{L_y}{l_{s\perp}}})\rightarrow{}\frac{l_s}{\sigma}\coth{\frac{L_y}{2l_s}}-2\frac{l_s}{\sigma}\coth\frac{L_y}{l_s} = -\frac{l_s}{\sigma}\tanh{\frac{L_y}{2l_s}}$, so that Eq. (\ref{eq:DeltarhogeneralSI}) gives
\begin{align}
    \Delta \rho_L &= \frac{2T_{xy}^2 l_{s}}{\sigma L_y}\tanh{\frac{L_y}{2l_{s}}}\frac{\sigma + |G_s|l_{s}\coth{\frac{L_y}{l_{s}}}-\bar{G}l_{s}\tanh{\frac{L_y}{2l_s}}\sin^2\phi}{\sigma  + |G_s|l_S \coth{\frac{L_y}{l_s}}} \\
    & = \frac{2T_{xy}^2 l_{s}}{\sigma L_y} \frac{\sigma \tanh{\frac{L_y}{2l_s}}+|G_s|l_s}{\sigma  + |G_s|l_S \coth{\frac{L_y}{l_s}}}-\frac{2T_{xy}^2 l_{s}}{\sigma L_y}\frac{\bar{G}l_s \tanh^2\frac{L_y}{2l_s}}{\sigma  + |G_s|l_S \coth{\frac{L_y}{l_s}}}\sin^2\phi\;.
\end{align}

Identifying $\sin^2\phi = 1-(\vec{m}\cdot\vec{N})^2$, and labelling the first term $\Delta\rho_0$ and the second term $\Delta\rho_1$, we recover Eqs. (\ref{eq:Longitudinal}-\ref{eq:Longitudinalrho1iso}) in the main text.

Next, consider the limit of highly anisotropic relaxation in which case $l_{s\perp}\rightarrow{}0$. so that also $l_{s\perp}\coth{\frac{L_y}{l_{s\perp}}}\rightarrow 0$
In this limit $\bar{G} = G_r$. Moreover, since in realistic junctions $|G_s|\ll G_r$, in the main text we set $G_s\rightarrow 0$

In that case Eq. (\ref{eq:DeltarhogeneralSI}) becomes

\begin{align}
    \Delta\rho_L = \frac{2T_{xy}^2 l_{s}}{\sigma L_y}\tanh{\frac{L_y}{2l_{s\parallel}}}\frac{\sigma + G_rl_{s\parallel}\coth{\frac{L_y}{2l_{s\parallel}}}\sin^2\phi}{\sigma + 2G_rl_{s\parallel}\coth{\frac{L_y}{l_{s\parallel}}}\sin^2\phi}\;.
\end{align}
Again identifying $\sin^2\phi = 1-(\vec{m}\cdot\vec{N})^2$, we obtain Eq. (\ref{eq:Largetausperp}) in the main text.

\bmsection{SMR in metallic antiferromagnets}\label{sec:SMRAFM}
To disentangle the effects of altermagnetism and spin-orbit coupling it is important to compare the SSMR with the SMR effect in antiferromagnets. To study the SMR effect in antiferromagnets, one actually has to resort to diffusion equations that incorporate both magnetism and spin-orbit coupling, which leads to a whole range of terms, such as relaxation and an anisotropic spin-Hall tensor.

However, to lowest order, we may consider the diffusion equations for an antiferromagnet \cite{kokkeler2025quantum}, which is the same as for altermagnets but with $T_{jk} = K_{jk} = 0$ due to the absence of spin-splitting, and add an isotropic spin-Hall tensor $\theta\epsilon_{jkl}$. Indeed, any other spin-to-charge conversion effects are of higher order in quasiclassics and therefore small.

We choose the current direction to be $\hat{x}$ and the normal along $y$. Since there are now three directions in spin space ($\vec{n}\times\vec{j},\vec{m},\vec{N}$) general expressions become rather expansive. We therefore decide to linearize in $G_{i,r,s}$ for illustrative purposes.

Like in the discussion of the influence of relaxation in altermagnets, we focus on two limits, relaxation due to magnetic impurities dominates over the Dyakonov-Perel type of mechanism, and the other way around.
\bmsubsection{Isotropic relaxation}
In the former case the spin-relaxation tensor is isotropic, $l_{s\parallel} = l_{s\perp} = l_s$. Since in the normal state magnetism-related relaxation is indistinguishable from relaxation due to SOC, the equations in this case are just those of the original SMR theory \cite{chen2013theory}. To lowest order in $G_{i,r,s}$, the spin accumulation reads
\begin{align}
    \vec{z}\cdot \vec{\mu}^s = 2\theta_{xyz}El_s\Big[\frac{\sinh\frac{y}{l_s}}{\cosh{\frac{L_y}{2l_s}}}+2\tanh{\frac{L_y}{2l_s}}\frac{\cosh\frac{y-\frac{L_y}{2}}{l_s}}{\sinh{\frac{L_y}{l_s}}}\Big(\frac{|G_s|l_s}{\sigma}+\frac{G_rl_s}{\sigma}(1-\vec{z}\cdot\vec{m})^2\Big)\Big]\;,
\end{align}
and the currents are given by
\begin{align}
    J_x &= L_y\sigma E -\frac{1}{2}\sigma \theta_{yxz}\int_{-\frac{L_y}{2}}^{\frac{L_y}{2}}dy\partial_y (\vec{z}\cdot \vec{\mu}^s)\nonumber\\
    &= L_y\sigma E + 2\theta_{xyz}^2\sigma E l_s\tanh{\frac{L_y}{2l_s}}-2\theta_{xyz}^2\sigma E \tanh^2\frac{L_y}{2l_s}\Big(\frac{|G_s|}{\sigma}+G_{r}(1-(\vec{z}\cdot\vec{m})^2)\Big)\;.
\end{align}
Note the sign difference compared to the SSMR case is purely due to $\theta_{yxz} = -\theta_{xyz}$.
\bmsubsection{Anisotropic relaxation}
Consider the case in which the relaxation for spins perpendicular to the axis is much stronger than for those parallel to it. In this case we obtain $\vec{\mu}^s\approx\mu^s \vec{N}$ and we have
\begin{align}
    \partial_{yy} \mu^s &= \frac{1}{l_{s\parallel}^2}\mu^s\;.\\
    \partial_y\mu^s &= 2\theta_{xyz}E\vec{N}\cdot\vec{z}\;,&y = \frac{L_y}{2}\;.\\
    \partial_y\mu^s & =2\theta_{xyz}E\vec{N}\cdot\vec{z} + 2(\frac{|G_s|}{\sigma}-\frac{G_r}{\sigma}\vec{N}\cdot(\vec{m}\times(\vec{m}\times \vec{N}))-G_i\vec{N}\cdot(\vec{m}\times\vec{N}))\mu^s&\nonumber\\& = 2\theta_{xyz}E\vec{N}\cdot\vec{z} + 2(\frac{|G_s|}{\sigma}+\frac{G_r}{\sigma}(1-(\vec{N}\cdot\vec{m})^2)\mu^s\;,&y = -\frac{L_y}{2}\;.
\end{align}
The solution to this equation is
\begin{align}
    \mu^s = \vec{N}\cdot\vec{z}2\theta_{xyz}El_{s\parallel}\Big[\frac{\sinh{\frac{y}{l_{s\parallel}}}}{\cosh{\frac{L_y}{2l_{s\parallel}}}}+2\tanh{\frac{L_y}{2l_{s\parallel}}}\frac{\cosh{\frac{y-\frac{L_y}{2}}{l_{s\parallel}}}}{\sinh{\frac{L_y}{l_{s\parallel}}}}\Big(\frac{|G_s|l_{s\parallel}+G_rl_{s\parallel}(1-(\vec{N}\cdot\vec{m})^2)}{\sigma+2(|G_s|l_{s\parallel}+G_rl_{s\parallel}(1-(\vec{N}\cdot\vec{m})^2))\coth{\frac{L_y}{l_{s\parallel}}}}\Big)\Big]\;.
\end{align}
With this, we have
\begin{align}
    J_x &= L_y\sigma E -\frac{1}{2}\sigma \theta_{yxz}\int_{-\frac{L_y}{2}}^{\frac{L_y}{2}}dy\partial_y (\vec{z}\cdot\vec{N}\mu^s)=L_y\sigma E + 2\theta_{xyz}^2\sigma E l_s\tanh{\frac{L_y}{2l_s}}(\vec{N}\cdot\vec{z})^2\nonumber\\
    &-2\theta_{xyz}^2\sigma(\vec{N}\cdot\vec{z})^2 E l_s \tanh^2\frac{L_y}{2l_s}\Big(\frac{|G_s|l_{s\parallel}+G_rl_{s\parallel}(1-(\vec{N}\cdot\vec{m})^2)}{\sigma+2(|G_s|l_{s\parallel}+G_rl_{s\parallel}(1-(\vec{N}\cdot\vec{m})^2))\coth{\frac{L_y}{l_{s\parallel}}}}\Big)\nonumber\\ & = L_y\sigma E + 2\theta_{xyz}^2\sigma E l_s\tanh{\frac{L_y}{2l_s}}(\vec{N}\cdot\vec{z})^2\frac{\sigma + (|G_s|l_{s\parallel}+G_rl_{s\parallel}(1-(\vec{N}\cdot\vec{m})^2))\coth{\frac{L_y}{2l_{s\parallel}}}}{\sigma+2(|G_s|l_{s\parallel}+G_rl_{s\parallel}(1-(\vec{N}\cdot\vec{m})^2))\coth{\frac{L_y}{l_{s\parallel}}}}\;.
\end{align}
Meanwhile the transverse current is
\begin{align}
    J_z &= \int_{-\frac{L_y}{2}}^{\frac{L_y}{2}}dy-\sigma(\partial_z\mu+\theta\epsilon_{yzx}\partial_y\mu^s\vec{N}\cdot\vec{x}+\theta\epsilon_{xzy}\partial_x\mu^s\vec{N}\cdot\vec{y})\nonumber\\&=\int _{-\frac{L_y}{2}}^{\frac{L_y}{2}}dy -\sigma\theta_{xyz}\vec{N}\cdot\vec{x}\partial_y\mu^s\nonumber\\&=\frac{\vec{N}\cdot\vec{x}}{\vec{N}\cdot\vec{z}}(J_x-L_y\sigma E)\;.
\end{align}
Therefore, the resistivity correction is
\begin{align}
    \Delta \rho_L &= -2\theta_{xyz}^2\frac{l_s}{L_y}\tanh{\frac{L_y}{2l_s}}(\vec{N}\cdot\vec{z})^2\frac{\sigma + (|G_s|l_{s\parallel}+G_rl_{s\parallel}(1-(\vec{N}\cdot\vec{m})^2))\coth{\frac{L_y}{2l_{s\parallel}}}}{\sigma+2(|G_s|l_{s\parallel}+G_rl_{s\parallel}(1-(\vec{N}\cdot\vec{m})^2))\coth{\frac{L_y}{l_{s\parallel}}}}\;,\\
    \Delta \rho_T &= \frac{\vec{N}\cdot\vec{x}}{\vec{N}\cdot\vec{z}}\Delta\rho_L = -2\theta_{xyz}^2\frac{l_s}{L_y}\tanh{\frac{L_y}{2l_s}}(\vec{N}\cdot\vec{z})(\vec{N}\cdot\vec{x})\frac{\sigma + (|G_s|l_{s\parallel}+G_rl_{s\parallel}(1-(\vec{N}\cdot\vec{m})^2))\coth{\frac{L_y}{2l_{s\parallel}}}}{\sigma+2(|G_s|l_{s\parallel}+G_rl_{s\parallel}(1-(\vec{N}\cdot\vec{m})^2))\coth{\frac{L_y}{l_{s\parallel}}}}\;.
\end{align}
We thus see that SMR in antiferromagnets has features of both the usual SMR and of the SSMR. Indeed, like the SSMR, the resistance depends on the angle between the magnetization of the FI and the N\'eel vector and the transverse and longitudinal resistivities have the same angle dependence. Moreover, like in SSMR, the anisotropic relaxation leads to an anharmonicity due to the appearance of $1-(\vec{N}\cdot\vec{m})^2$ not only in the numerator, but also in the denominator.

On the other hand, the sign of SMR for antiferromagnets is the same as for the SMR for normal metals, it is minimized along a specific axis and maximized along a plane. Moreover, this sign difference means that the anharmonicity leads to a widening of the peaks and tightening of the dips, opposite to the SSMR.  The magnitude of the effect is relativistic and it is determined by the relative direction of $\vec{N}$ and $\vec{z}$. This shows that the SMR in antiferromagnets can be distinguished both from the usual SMR and from the SSMR. 
In the case spin-relaxation is only slightly anisotropic the MR depends on both $\vec{m}\cdot\vec{N}$ and $\vec{m}\cdot\vec{z}$. This case can therefore be clearly distinguished from conventional SMR and SSMR as well.

\end{strip}

\end{document}